\documentclass[amsmath,amssymb,aps,pra,preprint,groupedaddress]{revtex4-1}

\usepackage[dvipdfmx]{graphicx}
\usepackage{dcolumn}
\usepackage{bm}
\usepackage{longtable}
\usepackage{ulem}
\usepackage{xcolor,soul}

\begin{document}

\title{Monte Carlo studies of skyrmion stabilization under geometric confinement and uniaxial strain}


\author{G. Diguet}
 \email{gildas.diguet.d4@tohoku.ac.jp}
\affiliation{%
Micro System Integration Center, Tohoku University, Sendai, Japan
}%
\author{B. Ducharne}
\affiliation{%
INSA Lyon, Universite de Lyon, Villeurbanne Cedex, France
}%
\affiliation{ElyTMax, CNRS-Universite de Lyon-Tohoku University, Sendai, Japan}
\author{S. El Hog}
\affiliation{%
Universit$\acute{\rm e}$ de Monastir (LMCN), Monastir, Tunisie
}
\author{F. Kato}%
\affiliation{
  National Institute of Technology (KOSEN), Ibaraki College, Hitachinaka, Japan.
}
\author{H. Koibuchi} 
\email{koibuchi@gm.ibaraki-ct.ac.jp; koibuchih@gmail.com}
\affiliation{
  National Institute of Technology (KOSEN), Ibaraki College, Hitachinaka, Japan.
}
\author{T. Uchimoto}
\affiliation{%
Institute of Fluid Science (IFS), Tohoku University, Sendai, Japan
}%
\affiliation{ELyTMaX, CNRS-Universite de Lyon-Tohoku University, Sendai, Japan}
\author{H. T. Diep}
 \email{diep@cyu.fr}
\affiliation{%
CY Cergy Paris University, Cergy-Pontoise, France
}%




\begin{abstract}
Geometric confinement (GC) of skyrmions in nanodomains plays a crucial role in skyrmion stabilization. This confinement effect decreases the magnetic field necessary for skyrmion formation and is closely related to the applied mechanical stresses. However, the mechanism of GC is unclear and remains controversial. Here, we numerically study the effect of GC on skyrmion stabilization and find that zero Dzyaloshinskii-Moriya interaction (DMI) coupling constants imposed on the boundary surfaces of small thin plates cause confinement effects, stabilizing skyrmions in the low-field region. Moreover, the confined skyrmions are further stabilized by tensile strains parallel to the plate, and the skyrmion phase extends to the low-temperature region. This stabilization occurs due to the bulk anisotropic DMI coupling constant caused by lattice deformations. Our simulation data are qualitatively consistent with  reported experimental data on skyrmion stabilization induced by tensile strains applied to a thin plate of the chiral magnet ${\rm Cu_2OSeO_3}$.
\end{abstract}

\maketitle


\section{Introduction \label{intro} }
Stabilization/destabilization of skyrmions \cite{Skyrme-1961,Moriya-1960,Dzyalo-1964} is a key target for future technological applications \cite{Uchida-etal-SCI2006,Yu-etal-Nature2010,Romming-etal-Science2013,Fert-etal-NatReview2017,Zhang-etal-JPhys2020,Gobel-etal-PhysRep2021}. The magnetic field $B$ plays a crucial role in skyrmion stabilization, and mechanical stresses and strains also strongly influence skyrmion stability \cite{Bogdanov-PRL2001,Butenko-etal-PRB2010}. Various experimental and theoretical studies have been conducted to identify the mechanisms of skyrmion stability \cite{Levatic-etal-SCRep2016,Pfleiderer-etal-Science2009,Yu-etal-PRB2015,Buhrandt-PRB2013}. Nii et al. reported that skyrmions in MnSi are stabilized (destabilized) by compressions perpendicular (parallel) to the magnetic field, improving the understanding of the skyrmion creation/annihilation mechanism \cite{Nii-etal-NatCom2015}. Charcon et al. reported that the area of the skyrmion phase in the $BT$ phase diagram increases or decreases depending on the compression direction, where $T$ is the temperature \cite{Charcon-etal-PRL2015}.

For the deformation of skyrmions by mechanical strains, Shibata et al. reported that skyrmions on thin FeGe films deform as oblong shapes along the direction of the tensile stress \cite{Shibata-etal-Natnanotech2015}. Mechanical stresses have been found to be significant in this phenomenon \cite{EWLee-RPP1955,Plumer-Walker-JPC1982,Plumer-etal-JPC1984,Kataoka-JPSJ1987}, and the shape deformation was successfully simulated with suitable magnetoelastic coupling terms \cite{Shi-Wang-PRB2018,Wang-Shi-Kamlah-PRB2018,Wang-ARMR2019}. In addition, skyrmion deformation was numerically obtained in two-dimensional simulations by assuming anisotropic Dzyaloshinskii-Moriya interaction (DMI) coefficients in Ref. \cite{Shibata-etal-Natnanotech2015}. This DMI anisotropy was also predicted based on a quantum mechanical mechanism \cite{Koretsune-etal-SCRep2015}. Moreover, the shape deformation phenomenon was studied with a $\bf{Z}_2$ vortex structure under antiferromagnetic coupling \cite{Osorio-etal-PRB2019} and was also simulated with the Finsler geometry modeling technique without assuming magnetoelastic coupling \cite{SElHog-etal-PRB2021,SElHog-etal-RIP2022}.

Anisotropic ferromagnetic coupling constants have also been shown to stabilize skyrmions. Anisotropy in the ferromagnetic interaction (FMI) arising from Rashba spin-orbit coupling enhances skyrmion stability on interfaces with inversion asymmetry \cite{Banerjee-etal-NPhys2013,Banerjee-etal-PRX2014}.
Chen et al. reported that FMI anisotropy in the easy axis enhances skyrmion stability in a 2D lattice model with isotropic DMI \cite{Chen-etal-Srep2016}, and enhanced stability was observed in a 3D lattice model with both FMI and DMI anisotropy, indicating uniaxial stress effects \cite{Chen-etal-Srep2017}. Strain-induced stabilization was simulated by assuming anisotropic DMI constants \cite{Tanaka-etal-PRM2020}.
FMI and DMI anisotropy inducing uniaxial stress effects increases the area of the skyrmion phase in $BT$ phase diagrams \cite {WCLi-etal-PhysScr2022}, and anisotropy in antisymmetric FMIs effectively produces a chiral magnetic interaction corresponding to DMIs \cite{SGao-etal-Nat2020,DAmoroso-etal-Nat2020}.
These anisotropic FMIs are important in obtaining the domain wall width of layered two-dimensional magnetic materials \cite{HHYang-etal-2DMat2022} that exhibit the so-called nonreciprocal propagation of magnons on the surface \cite{MCosta-etal-PRB2020}. Antisymmetric FMIs between thin layered materials effectively induce FMI anisotropy and have been shown to reduce the skyrmion Hall effect, and consequently, anisotropic FMIs enhance the transport stability in thin linear domains \cite{Zhang-etal-Natcom2016,Mukherjee-etal-PRB2021}.

Another stabilization mechanism is the geometric confinement (GC) effect studied in Ref.\cite{Rohart-Thiaville-PRB2013}, which assumes magnetization anisotropy and a constant DMI coefficient. A GC effect was experimentally observed in a FeGe nanostripe \cite{HDu-etal-NatCom2015}, and morphological changes in skyrmions with varying nanostripe widths were reported in \cite{CJin-etal-NatCom2017}. Skyrmion bubbles in centrosymmetric magnets are also influenced by GC effects \cite{ZHou-etal-AcsNano2019}, where the applied magnetic field decreases with decreasing nanostripe width, indicating stabilization. Furthermore, Ho et al. reported that confined skyrmions are stabilized in multilayered nanodomains \cite{PHo-etal-PRAp2019}.

\begin{figure}[h]
\centering{}\includegraphics[width=10.5cm,clip]{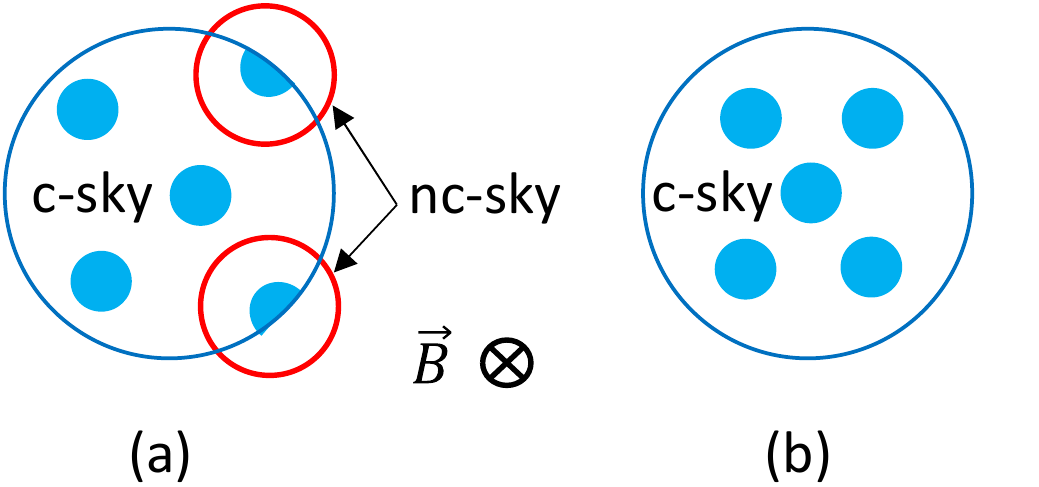}
\caption{\label{fig-1}
Illustrations of a (a) nonconfined skyrmion (nc-sky) configuration and a (b) confined skyrmion (c-sky) configuration in a small disk domain. The red circles in (a) enclose nc-sky and unstable skyrmions touching the boundary. The external magnetic field $\vec{B}$ is applied perpendicular to the disk.
}
\end{figure}
Recently, Wang et al. reported experimental data on the switching mechanism for individual skyrmions in nanodots with diameters ranging from $150 ({\rm nm})$ to $1000 ({\rm nm})$ \cite{YWang-etal-NatCom2020}. In their study, a magnetic field $\vec{B}$ was applied perpendicular to the disk, and a variable tensile strain was electrically applied in the radial direction via a substrate. The reported data show that skyrmions are confined in the nanodots and that $\vec{B}$ decreases with decreasing nanodot diameter.
This reduction in $\vec{B}$ is expected to be a consequence of both GC and magnetoelastic effects \cite{YWang-etal-NatCom2020}. Figs. \ref{fig-1}(a) and (b) illustrate nonconfined and confined skyrmions in a small disk domain. In small domains, surface effects are expected to be strong, and no nonconfined skyrmions were reported in Ref. \cite{YWang-etal-NatCom2020}. The main target of the study in Ref. \cite{YWang-etal-NatCom2020} was not the GC effect but rather electric field-driven switching among individual skyrmions; however, the results indicate that GC is closely connected to this switching.

Seki et al. reported experimental data on the dependence of the direction of the magnetic field $\vec{B}$ on a small thin plate of the chiral magnet ${\rm Cu_2OSeO_3}$, where the thickness of the specimen is $1({\rm \mu m})$ \cite{Seki-etal-PRB2017}. The reported data show remarkable stabilization when $\vec{B}$ is perpendicular to the strain direction and parallel to the plate surface. We should note that ${\rm Cu_2OSeO_3}$ is stabilized by extensions perpendicular to $\vec{B}$, while MnSi in Refs. \cite{Nii-etal-NatCom2015,Charcon-etal-PRL2015} is stabilized by compressions perpendicular to $\vec{B}$. These observations indicate that the response of ${\rm Cu_2OSeO_3}$ differs from those of MnSi and FeGe, at least in the case of mechanical strain, because FeGe in Ref. \cite{Shibata-etal-Natnanotech2015} is expected to be destabilized by compressions parallel to $\vec{B}$. However, this stabilization enhancement in ${\rm Cu_2OSeO_3}$ is not indicated by the combined effects of strains and GC.

In this paper, we perform Monte Carlo simulations of the GC effect for skyrmions in a 3D lattice discretized by tetrahedra, carefully investigating the effects of DMI coefficients on skyrmion stabilization. In the simulation model, we assume DMI coefficients of zero on the boundary surfaces parallel to the magnetic field by modifying the DMI vector $\vec{D}_{ij}$ in the DMI energy term $\sum_{ij} \vec{D}_{ij}\cdot \vec{\sigma}_i\!\times\!\vec{\sigma}_j$, where $\vec{\sigma}_i(\in S^2\;{\rm unit\; sphere})$ is the spin variable, and show that this assumption notably improves skyrmion stability. The effects of strains on the stabilization are also investigated by assuming lattice deformation corresponding to tensile deformation without the magnetoelastic coupling terms in the Hamiltonian. Thus, in our model, the DMI vector is modified heterogeneously by the GC effect and anisotropically by uniaxial strains. Specifically, Bloch-type skyrmions are studied in this paper: $\vec{D}_{ij}\!=\!\vec{e}_{ij}$, where $\vec{e}_{ij}$ is a tangential vector from spin positions $i$ to $j$. To develop a model considering GC and strain effects, we carefully analyze the results in Ref. \cite{Koibuchi-etal-ICMsquare2022} for Neel-type skyrmions, which are defined as $\vec{D}_{ij}\!=\!\vec{e}_{ij}\!\times\!\vec{e}^{\;z}_{i}$ \cite{ERuff-etal-SciAdv2015,IKezsmarki-etal-NatMat2015,YFujimka-etal-PRB2017,YWu-etal-NatCom2020}, where $\vec{e}^{\;z}_{i}$ denotes the magnetic field direction, in Section \ref{background}.

\noindent
\section{Results of Neel-type skyrmion models \label{background} }
In this section, we briefly review the simulation results in Ref. \cite{Koibuchi-etal-ICMsquare2022}, which does not provide detailed information about the models and confinement mechanism. After a short review of the simulation results and the models, we emphasize that the position dependence of the DMIs in the model in Ref. \cite{Koibuchi-etal-ICMsquare2022} gives us a crucial hint for defining the geometric confinement model, which is introduced in the following section.

\subsection{Dzyaloshinskii-Moriya interaction-dependent confinement \label{DMI-dependent-conf}}
\begin{figure}[h]
\centering{}\includegraphics[width=11.5cm,clip]{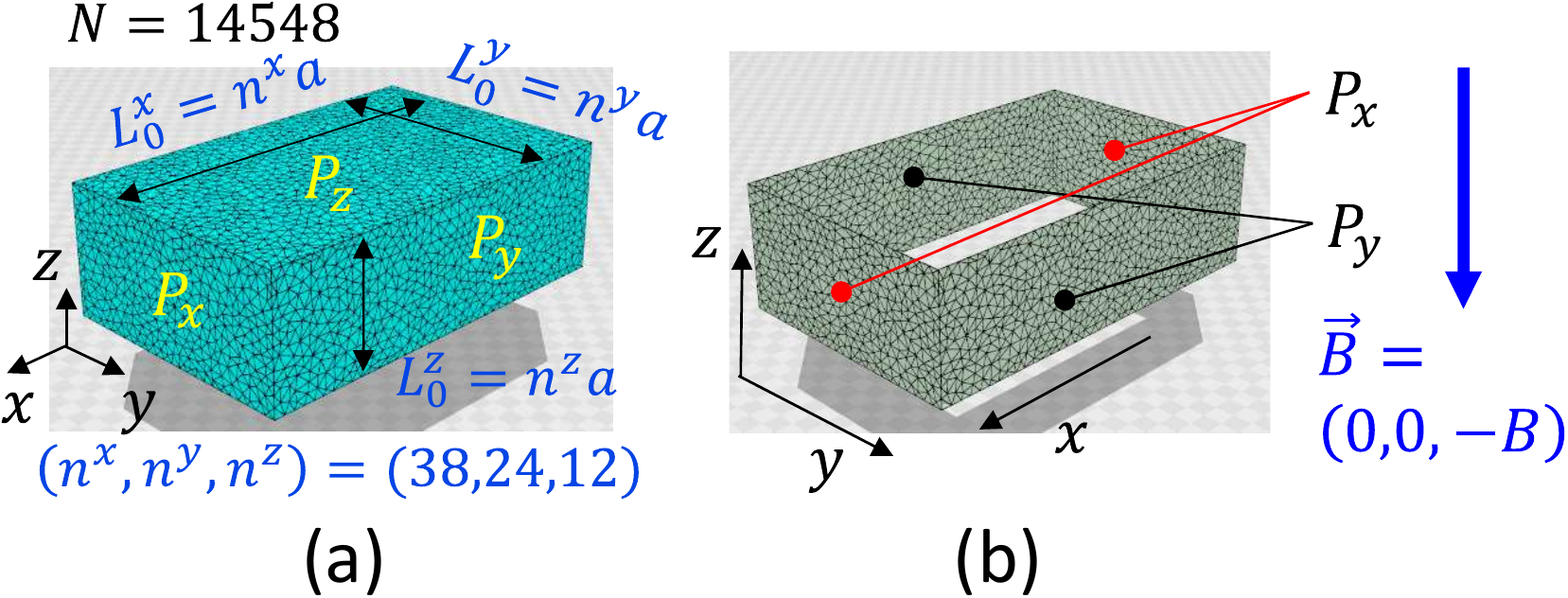}
\caption{(a) 3D lattice discretized by tetrahedra (see Appendix \ref{App-A} for information on the lattice construction), where the total number of vertices is $N\!=\!14548$. (b) Boundary surfaces, denoted by $P_x$ and $P_y$, parallel to the magnetic field direction $\vec{B}\!=\!(0, 0, -B)$. The boundary surface $P_\mu$ is defined by $P_\mu\perp \vec{e}_{\mu}$, where $\vec{e}_{\mu}, (\mu=x,y,z)$ is the unit vector along the $\mu$ direction. This lattice is also used in the following section.
\label{fig-2} }
\end{figure}

In Ref. \cite{Koibuchi-etal-ICMsquare2022}, Metropolis Monte Carlo (MMC) simulations \cite{Metropolis-JCP-1953,Landau-PRB1976} were performed to update the spin variables $\vec{\sigma}$ on a three-dimensional (3D) cubic lattice (Fig. \ref{fig-2}(a)) under free boundary conditions, with the magnetic field applied along the $z$ direction, as shown in Fig. \ref{fig-2}(b). In the MC update $\vec{\sigma}_i\!\to\! \vec{\sigma}_i^\prime$ at the lattice site, the new variable $\vec{\sigma}_i^\prime$ is randomly distributed on the unit sphere $S^2$ independent of the original variable $\vec{\sigma}_i$, and $\vec{\sigma}_i^\prime$ is accepted with probability ${\rm Max}[1,\exp({\delta}S/T)]$, where ${\delta}S\!=\!S(\cdots,\vec{\sigma}_i^\prime,\cdots)\!-\!S(\cdots,\vec{\sigma}_i,\cdots)$ and $T$ is the temperature. In this expression, $S$ is the total Hamiltonian, which is shown below. The ground state was assumed for the initial configurations of $\vec{\sigma}$ in these MMC simulations. The technique for finding the ground state is described below. The lattice size is given by $(L^x_0,L^y_0,L^z_0)\!=\!(n^xa,n^ya,n^za)\!=\!(38a,24a,12a)$, where $a$ is a length unit known as the lattice spacing. The lattice spacing is isotropic and corresponds to the mean edge length of the tetrahedra. The mean edge length corresponds to the mean distance between two neighboring atoms in a coarse-grained approach or groups of atoms, as in other lattice models \cite{Creutz-txt}.

\begin{figure}[h]
\centering{}\includegraphics[width=11.5cm,clip]{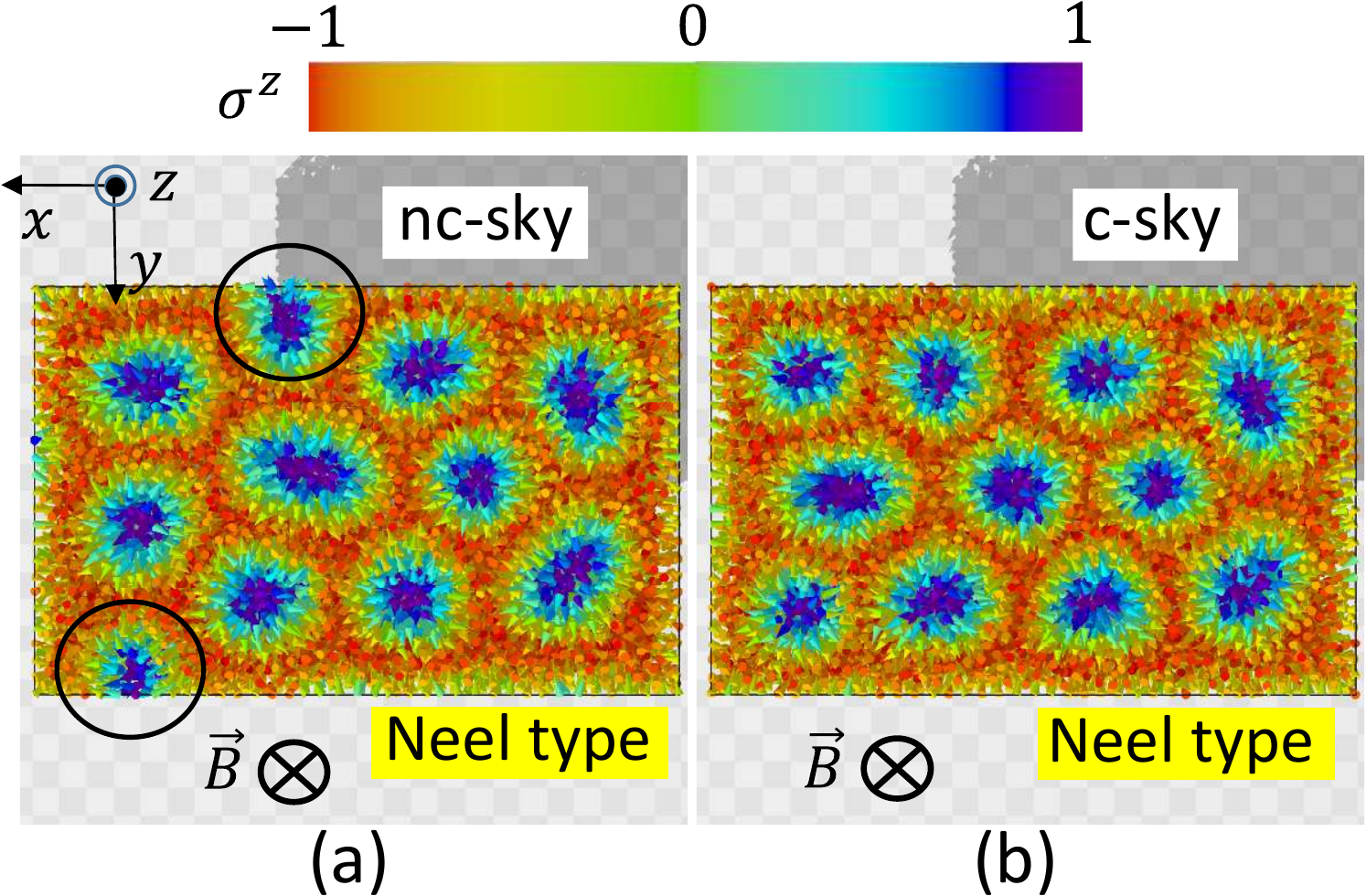}
\caption{Snapshots of Neel-type skyrmion configurations with (a) nonconfined (nc-sky) and (b) confined (c-sky) skyrmions. The skyrmions in (a) enclosed by the solid circles touching the boundary $P_y$ (Fig. \ref{fig-2}(b)) are reported to be unstable, and all skyrmions in (b) confined inside the boundary are stable \cite{Koibuchi-etal-ICMsquare2022}. The color legends correspond to the $\sigma^z$ of the spins drawn by small cones.
\label{fig-3} }
\end{figure}

We next present snapshots of the simulation results for the two different types of skyrmion configurations reported in Ref. \cite{Koibuchi-etal-ICMsquare2022}, namely, confined skyrmions (c-sky) and nonconfined skyrmions (nc-sky), as shown in Figs. \ref{fig-3}(a) and (b). The skyrmions enclosed by the black circles in Fig. \ref{fig-3}(a) were reported to be unstable; some of the skyrmions touching the boundary disappear, and new skyrmions emerge on the boundary after long MC simulations \cite{Koibuchi-etal-ICMsquare2022}. As a result, the positions of the nonconfined skyrmions may fluctuate or change, while the positions of the confined skyrmions remain unchanged.

The Hamiltonians introduced in \cite{Koibuchi-etal-ICMsquare2022} corresponding to these configurations are given by
\begin{eqnarray}
 \label{total-Hamiltonian}
 \begin{split}
&S=\lambda S_{{\rm FM}}+DS_{{\rm DM}}-S_{B},\\
&S_{B}=\sum_{i}\vec{\sigma}_{i}\cdot\vec{B},\quad\vec{B}=(0,0,-B), 
\end{split}
\end{eqnarray}
where the Zeeman energy $S_{B}$ has the same expression in the two models corresponding to the snapshots shown in Figs. \ref{fig-3}(a) and (b).
The symbol $\vec{B}$ denotes the external magnetic field.
The models differ with regard to their DMI energy $S_{\rm DM}$ and FMI energy $S_{\rm FM}$, which are
given by
\begin{eqnarray}
&\left\{ \begin{array}{@{\,}ll}
                 S_{\rm FM}=\sum_{ij}\left(1-\vec{\sigma}_{i}\cdot\vec{\sigma}_{j}\right), \\
                 S_{{\rm DM}}=\sum_{ij}({\vec{e}}_{ij}\times \vec{e}_z)\cdot(\vec{\sigma}_{i}\times\vec{\sigma}_{j}), 
                  \end{array} 
                   \right.    \quad   ({\rm model \; for \; Fig. \ref{fig-3}(a)}),
\label{model-Fig-AB-1} \\
&\left\{ \begin{array}{@{\,}ll}
                 S_{\rm FM}=\sum_{ij}n_{ij}\left(1-\vec{\sigma}_{i}\cdot\vec{\sigma}_{j}\right), \\
                 S_{{\rm DM}}=\sum_{ij}n_{ij}({\vec{e}}_{ij}\times  \vec{e}_z)\cdot(\vec{\sigma}_{i}\times\vec{\sigma}_{j}),
                  \end{array} 
                   \right.    \quad   ({\rm model \; for \; Fig. \ref{fig-3}(b)}),
\label{model-Fig-AB-2} \\
&\vec{e}_{ij}=\left(e_{ij}^x,e_{ij}^y,e_{ij}^z\right), \quad \|\vec{e}_{ij}\|=1, 
\label{model-Fig-AB-3}
\end{eqnarray}
where $\vec{e}_{ij}$ is the unit vector from vertices $i$ to $j$, the vector $\vec{e}_z(=\!(0, 0, 1))$ indicates the $\vec{B}$ direction, and $n_{ij}$ corresponds to the total number of tetrahedra sharing bond $ij$ with a normalization factor.
The factor $n_{ij}$ in the model shown in Fig. \ref{fig-3}(b) appears because the discretization assumed for these $S_{\rm FM}$ and $S_{\rm DM}$ in Ref. \cite{Koibuchi-etal-ICMsquare2022} is slightly different from the standard discretization technique corresponding to the standard Hamiltonian, such as $S_{\rm FM}\!=\!\sum_{ij}\left(1-\vec{\sigma}_{i}\cdot\vec{\sigma}_{j}\right)$. We should note that such a discrete Hamiltonian can be obtained with the assumed discretization scheme from the continuous Hamiltonian defined by using differentials and integrals. Therefore, in general, the discrete form of the Hamiltonian depends on the discretization scheme.

The Hamiltonians
in Eqs. (\ref{model-Fig-AB-1}) and (\ref{model-Fig-AB-2}) correspond to Neel-type skyrmions \cite{ERuff-etal-SciAdv2015,IKezsmarki-etal-NatMat2015,YFujimka-etal-PRB2017,YWu-etal-NatCom2020}, as mentioned in the Introduction. The two models defined by Eqs. (\ref{model-Fig-AB-1}) and (\ref{model-Fig-AB-2}) correspond to models 1 and 2, respectively, in \cite{Koibuchi-etal-ICMsquare2022}.

We emphasize that the skyrmions are confined by using the model of $S_{\rm FM}$ and $S_{\rm DM}$ shown in Eq. (\ref{model-Fig-AB-2}), where the $n_{ij}$ value on the surface is smaller than that on the inside. Thus, we consider that this difference in the DMI between the surface and bulk is closely connected to the confinement mechanism. In this sense, the DMI of the model formulated in Eq. (\ref{model-Fig-AB-2}) is position dependent.

\noindent
\section{Geometric confinement model\label{GC-model}}
This and the next sections are the main part of this paper.
In the previous section, we confirmed that skyrmions are confined in small domains if the surface DMI coefficient is substantially smaller than the bulk DMI coefficient. If the DMI coefficient is small on surfaces such as $P_x$ and $P_y$ in Fig. \ref{fig-2}(b), skyrmions cannot appear on $P_x$ and $P_y$ and are thus confined inside the domain boundary.
Based on this observation, in this paper, we propose a model in which the DMI coefficient is set to zero on the boundary surfaces parallel to the magnetic field $\vec{B}$, which is applied along the $y$ direction, as shown in Fig. \ref{fig-4}(a). The $\vec{B}$ direction is changed to study confinement effects in narrow domains such as nanostripes \cite{HDu-etal-NatCom2015,CJin-etal-NatCom2017}. The thickness $L^z_0\!=\!12$ of the lattice in the simulation unit ($a\!=\!1$) is sufficiently thin compared with the skyrmion size, allowing skyrmions to appear in the central region between the surfaces $P_z$, while $L^x_0\!=\!38$ and $L^y_0\!=\!24$ are sufficiently large compared with $L^z_0$ (Appendix \ref{App-A}).
\subsection{Hamiltonian and lattice deformation\label{model-GC}}
We emphasize that the large difference in the models presented in this and the preceding section originates from $n_{ij}$ in $S_{\rm DM}$ in Eq. (\ref{model-Fig-AB-2}). In this section, to evaluate surface effects in a GC model, we simply fix the DMI coefficient to zero on the boundary surfaces $P_x$ and $P_z$ in the standard discrete Hamiltonian $S_{\rm DM}$ instead of using $S_{\rm DM}$ in Eq. (\ref{model-Fig-AB-2}). As emphasized in the preceding section, this replacement of $S_{\rm DM}$ is motivated by the difference in the Hamiltonian discretization schemes.

\begin{figure}[h]
\centering{}\includegraphics[width=10.5cm,clip]{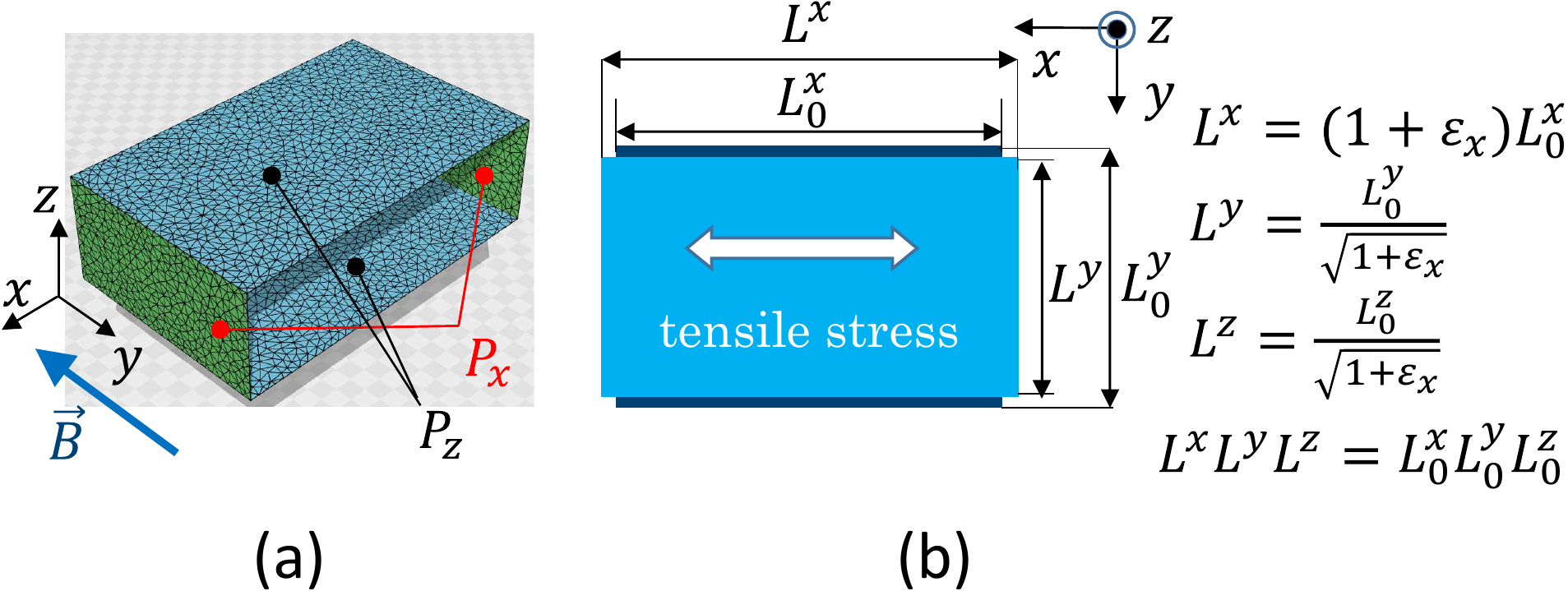}
\caption{(a) The DMI coefficients are fixed at zero on the boundary surfaces $P_x$ and $P_z$ parallel to the magnetic field $\vec{B}$ along the $y$ direction, and (b) lattice deformation characterized by $\varepsilon_x$ is induced by tensile stress along the $x$ direction, where $\varepsilon_x\!=\!0$ and $\varepsilon_x\!=\!0.02$ are assumed in the simulations, and the volume remains unchanged such that $L^xL^yL^z (\varepsilon_x\!=\!0.02)\!=\!L^x_0L^y_0L^z_0 (\varepsilon_x\!=\!0)$ for simplicity.
\label{fig-4}  }
\end{figure}

To observe the effect of a zero DMI coefficient on the boundary surface and to verify that only the zero DMI model shows GC effects, we study a standard model with $S_{\rm DM}$ defined on the bonds and no DMI position dependence. For the same reason, we also study a model in which the FMI coupling constant is fixed at zero on the surfaces parallel to $\vec{B}$. Thus, three different models, namely, model A, model B and model C, are studied in this paper. Model A is the standard model, model B is defined by zero DMI coefficients on $P_x\cup P_z$ to evaluate GC effects, and model C is defined by zero FMI on $P_x\cup P_z$:
\begin{eqnarray}
\label{difference-models-AB}
\begin{split}
&{\rm model \; A \quad (standard\; model)},\\
&{\rm model \; B \quad (model\; for\; geometric\; confinement}), \\
& \hspace{1.9cm} : {\rm zero\; DMI\; on\;} P_x\cup P_z,\\
&{\rm model \; C \quad : zero\; FMI\; on\;} P_x\cup P_z.
\end{split}
\end{eqnarray}

The total Hamiltonian $S$, which is the same as that in Eq. (\ref{total-Hamiltonian}), and the energy terms are defined as follows:
\begin{eqnarray}
\label{models-AB}
\begin{split}
&S=\lambda S_{{\rm FM}}+DS_{{\rm DM}}-S_{B},\\
& S_{B}=\sum_{i}\sigma_{i}\cdot\vec{B},\quad\vec{B}=(0,-B,0), 
\end{split}
\end{eqnarray}
\begin{eqnarray}
\label{model-A}
\begin{split}
&\left\{ \begin{array}{@{\,}ll}
                 S_{\rm FM}=\sum_{ij}\left(1-\vec{\sigma}_{i}\cdot\vec{\sigma}_{j}\right), \\
                 S_{{\rm DM}}=\sum_{ij} \vec{e}_{ij} \cdot(\vec{\sigma}_{i}\times\vec{\sigma}_{j}),
                  \end{array} 
                   \right.    \quad   ({\rm model \; A}), 
\end{split}
\end{eqnarray}
\begin{eqnarray}
\label{model-B}
\begin{split}
&\left\{ \begin{array}{@{\,}ll}
                 S_{\rm FM}=\sum_{ij}\left(1-\vec{\sigma}_{i}\cdot\vec{\sigma}_{j}\right), \\
                 S_{{\rm DM}}=\sum_{ij} \Gamma_{ij}{\vec{e}}^{\;\prime}_{ij}\cdot(\vec{\sigma}_{i} \times\vec{\sigma}_{j}),
                  \end{array} 
                   \right.    \quad   ({\rm model \; B\; for\; GC}),
\end{split}
\end{eqnarray}
\begin{eqnarray}
\label{model-C}
\begin{split}
&\left\{ \begin{array}{@{\,}ll}
                 S_{\rm FM}=\sum_{ij}\Gamma_{ij}\left(1-\vec{\sigma}_{i}\cdot\vec{\sigma}_{j}\right), \\
                 S_{{\rm DM}}=\sum_{ij} {\vec{e}}_{ij}\cdot(\vec{\sigma}_{i} \times\vec{\sigma}_{j}),
                  \end{array} 
                   \right.    \quad   ({\rm model \; C}),
\end{split}
\end{eqnarray}
\begin{eqnarray}
\label{for-model-BC}
\begin{split}
& \Gamma_{ij}=\left\{ \begin{array}{@{\,}ll}
                 0 &  (ij \in P_x\cup P_z) \\
                 1&  ({\rm otherwise}) 
                  \end{array} 
                   \right.,    \\
& \vec{e}^{\;\prime}_{ij}=\left((1+\varepsilon_x) e_{ij}^x, \frac{e_{ij}^y}{\sqrt{1+\varepsilon_x}},\frac{e_{ij}^z}{\sqrt{1+\varepsilon_x}}\right),\\
& 
\left(\|\vec{e}^{\;\prime}_{ij}\|\geq\|\vec{e}_{ij}\|=1\;{\rm for}\; \varepsilon_x\geq 0 \right).
\end{split}
\end{eqnarray}
The DMI energy $S_{\rm DM}$ of model B differs from that in models A and C. In model B, the surface condition $\Gamma_{ij}\!=\!0$ (on $P_x\cup P_z)$ and  $\Gamma_{ij}\!=\!1$ (otherwise) in Eq. (\ref{for-model-BC}) assumed in $S_{\rm DM}$ confines skyrmions. In addition to this confinement mechanism, to induce uniaxial strain effects in model B, we replace the unit vector $\vec{e}_{ij}$ along bond $ij$ with $\vec{e}^{\;\prime}_{ij}$. This
 $\vec{e}^{\;\prime}_{ij}$ represents a new direction of bond $ij$ that is neither parallel to $\vec{e}_{ij}$ nor of unit length when $\varepsilon_x\!\not=\! 0$. $\vec{e}^{\;\prime}_{ij}$ in Eq. (\ref{for-model-BC}) is adopted because such a modification of the DMI vector $\vec{D}_{ij}$ is expected during lattice deformation, which is discussed below. Moreover,$\vec{e}^{\;\prime}_{ij}$ originally corresponds to a tangential vector $\partial \vec {r}/\partial x$ along bond $ij$, and $\partial \vec {r}/\partial x$ is not always of unit length in general. 
$\Gamma_{ij}$ is included in $S_{\rm FM}$ in model C to show that the surface effects induced by the FMI does not lead to confinement and to confirm that the surface effects of the DMI confines only skyrmions. Here, we emphasize that strain effects on the FMI play prominent roles in skyrmion stabilization, as emphasized by reviewing previous studies in the Introduction. However, this topic is beyond the scope of this paper; we assume that only DMI deformation induced by the surface and strain effects causes skyrmion stabilization. The $S_{\rm DM}$ expressions in models A, B and C are of the Bloch type, in contrast to the cases defined in Eqs. (\ref{model-Fig-AB-1}) and (\ref{model-Fig-AB-2}) for Figs. \ref{fig-3}(a) and (b). The Bloch type is assumed here because the lattice thickness $L^y_0(=\!24a)$ along the $\vec{B}$ direction is not so small when compared with $L^z_0(=\!12a)$ in the case of the preceding section, as mentioned above.

The partition function $Z$ is given by
\begin{eqnarray}
\label{part-function}
Z=\sum _{\vec{\sigma}}\exp(-S(\vec{\sigma})/T),
\end{eqnarray}
where $\sum_{\vec{\sigma}}$ denotes the sum over all possible spin configurations $\vec{\sigma}\!=\!\{\vec{\sigma}_1,\vec{\sigma}_2,\cdots,\vec{\sigma}_N\}$. The simulation unit is given by $k_B\!=\!1$ and $a\!=\!1$, where $k_B$ and $a$ are the Boltzmann constant and the lattice spacing, respectively.

A tensile strain $\varepsilon_x(\geq 0)$ is applied along the $x$ axis of the lattice in model B 
 to examine the skyrmion stability in the low $T$ region expected from reported experimental data on the insulator ${\rm Cu_2OSeO_3}$ \cite{Seki-etal-PRB2017}, as mentioned in the Introduction. This strain $\varepsilon_x(\geq 0)$ deforms the lattice size as 
\begin{eqnarray}
\label{lattice-deform}
(L^x_0, L^y_0, L^z_0)\to (L^x, L^y, L^z)=\left((1+\varepsilon_x) L^x_0, \frac{L^y_0}{\sqrt{1+\varepsilon_x}},\frac{L^z_0}{\sqrt{1+\varepsilon_x}}\right), \quad (\varepsilon_x \ge 0).
\end{eqnarray}
This deformation condition ensures that the lattice volume remains unchanged, as shown in Fig. \ref{fig-4}(b), and explains why $\vec{e}^{\;\prime}_{ij}$ in Eq. (\ref{for-model-BC}) represents the direction of the bond $ij$ in the deformed lattice. Moreover, according to Eq. (\ref{lattice-deform}), the lattice spacing $a$ changes from isotropic to direction-dependent, such that $(a^x,a^y,a^z)\!=\!((1\!+\!\varepsilon_x)a, a/\sqrt{1\!+\!\varepsilon_x}, a/\sqrt{1\!+\!\varepsilon_x})$, because $L^\mu_0$ on the right-hand side is given by $L^\mu_0\!=\!n^\mu a$ (Fig. \ref{fig-2}(a) and Appendix \ref{App-A}).
However, the abovementioned condition $a\!=\!1$ is satisfied up to the order of $O(\varepsilon_x^3)$. Therefore, the simulation unit remains essentially unchanged for small $\varepsilon_x$ values, such as $\varepsilon_x\!=\!0.02$ assumed in the simulations. Under this condition in Eq. (\ref{lattice-deform}), the tensile stress along the $x$ axis is equivalent to the compressive stresses along the $y$ and $z$ axes, as discussed in Ref. \cite{SElHog-etal-RIP2022}. We note that magnetoelastic terms are not included in $S$; instead, the DMI coefficients effectively become direction- and position-dependent due to the surface effects caused by $\Gamma_{ij}$ and strain effects caused by the lattice deformation in Eq. (\ref{lattice-deform}). Detailed information regarding the effective DMI coefficients is provided in the following subsection.

\subsection{Effective coupling constant for the geometric confinement model}
\begin{figure}[h!]
\centering{}\includegraphics[width=10.5cm,clip]{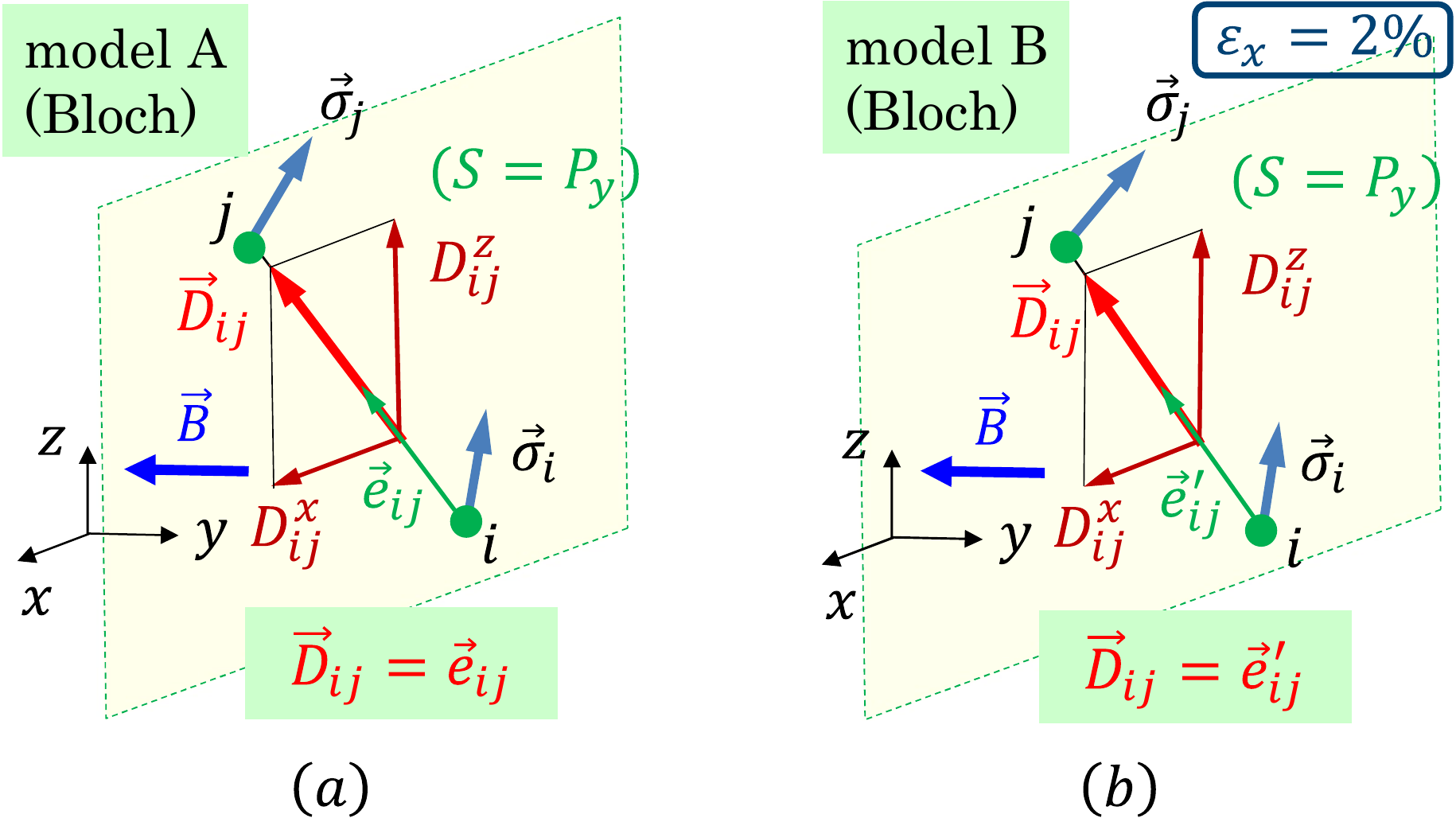}

\caption{(a) DMI vector $\vec{D}_{ij}\!=\!\vec{e}_{ij}$ of
model A, which is the standard model, on the surface $S\!=\!P_y$, and (b) DMI vector $\vec{D}_{ij}\!=\!\Gamma_{ij}\vec{e}^{\;\prime}_{ij}\!=\!\vec{e}^{\;\prime}_{ij}$ on $P_y$ of model B ($\varepsilon_x\!>\!0$), which is a model of geometric confinement with uniaxial strains. $\vec{\sigma}_i$ and $\vec{\sigma}_j$ are both on $P_y$ in (a) and (b). The $y$ axis component is $D^y_{ij}\!=\!0$ on $P_y$ in both models A and B because $\vec{e}_{ij}$ and $\vec{e}^{\;\prime}_{ij}$ are on $P_y$ in (a) and (b). Note that the lattice shapes differ because of the strain in model B. \label{fig-5}  }
\end{figure}
To reveal the origin of these morphological changes, namely, strain-induced stabilization, which is presented in the next section, we define the surface DMI such that
\begin{eqnarray}
\label{effective-D-SV-strain-A}
D^\mu_S=\frac{1}{\sum_{ij\in S}1}\sum_{ij\in S} |{e}_{ij}^{\mu}| \;\;{\rm on}\;S=(P_x\cup P_y\cup P_z)\setminus P_\mu,\;(\mu=x,y,z) \quad ({\rm model\; A}), 
\end{eqnarray}
where $S\!=\!(P_x\!\cup\! P_y\!\cup\! P_z)\setminus P_x$ means $S\!=\!P_y\!\cup\! P_z$, for example.
In model B, all $D^\mu_S (\mu\!=\!x, y, z)$ are defined to be zero on $P_x \cup P_z$, while $D^\mu_S (\mu\!=\!x, z)$ are nonzero on $P_y$. Therefore, we have
\begin{eqnarray}
\label{effective-D-SV-strain-B}
D^\mu_S=\left\{ \begin{array}{@{\,}ll}
                 0\;\;{\rm on} \;P_x\cup P_z,\; (\mu=x,y,z)  &  \\
                 \frac{1}{\sum_{ij\in P_y}1}\sum_{ij\in P_y} |{e}_{ij}^{\prime\;\mu}| \;\;{\rm on}\;P_y, \; (\mu=x,z) \; & 
                  \end{array} 
                   \right.,\quad ({\rm model\; B}),
\end{eqnarray}
where $D^y_S\!=\!0$ on $P_y$. The DMI vector $\vec{D}_{ij}\!=\!\vec{e}_{ij}$ of model A is the same as that of model C, and therefore, the corresponding constants $D^\mu_{S,V}$ are also common to models A and C. For this reason, we discuss the constants $D^\mu_{S,V}$ of only models A and B to simplify the notations in this subsection.

\begin{figure}[h!]
\centering{}\includegraphics[width=10.5cm,clip]{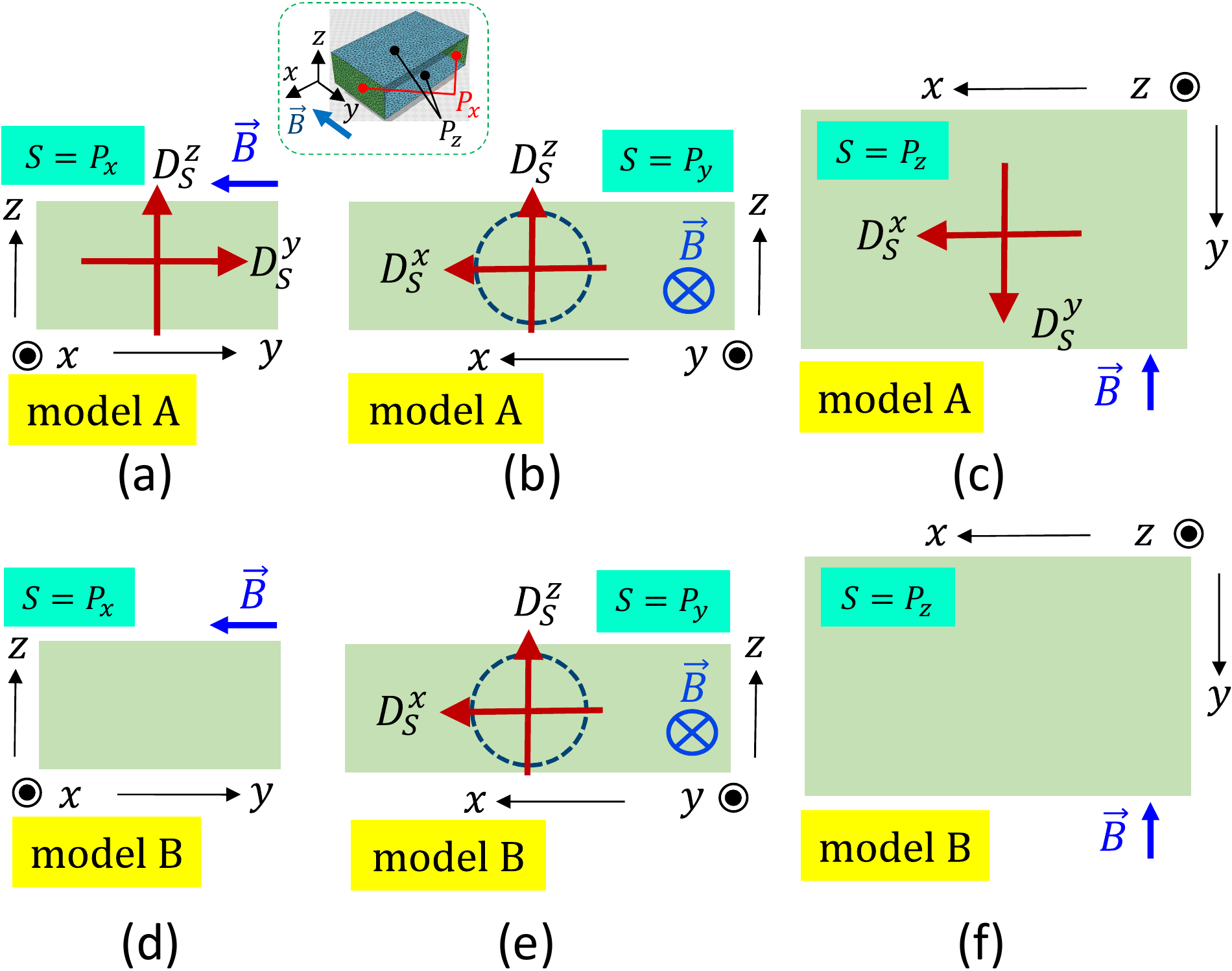}
\caption{Nonzero effective coupling constants $D^\mu_S\; (\mu\!=\!x,y,z)$ of model A on (a) $S\!=\!P_x$, (b) $S\!=\!P_y$, and (c) $S\!=\!P_z$, and $D^\mu_S$ of model B on (d) $S\!=\!P_x$, (e) $S\!=\!P_y$, and (f) $S\!=\!P_z$. The symbol $\otimes$ on $P_y$ in (b) and (e) denotes the $\vec{B}$ direction, and the dashed circle denotes an expected skyrmion configuration. In both models A and B, one component $D^\mu_S$ satisfying $D^\mu_S\!\perp\!S$ is $D^\mu_S\!=\!0$ because the DMI vectors $\vec{D}_{ij}\!=\!\vec{e}_{ij}$ and $\vec{D}_{ij}\!=\!\vec{e}^{\;\prime}_{ij}$ are on $S(=\!P_x,P_y,P_z)$, as indicated in Figs. \ref{fig-5}(a) and (b). In model B, $D^\mu_S\; \!=\!0, (\mu\!=\!x,y,z)$ on $P_x$ and $P_z$ according to the definition in Eq. (\ref{effective-D-SV-strain-B}).
\label{fig-6}  }
\end{figure}

The effective coupling constants $D^\mu_{S,V}$ are the mean component lengths of the DMI vectors $\vec{D}_{ij}\!=\!\vec{e}_{ij}$ for model A and $\vec{D}_{ij}\!=\!\vec{e}^{\;\prime}_{ij}$ for model B. The vectors $\vec{D}_{ij}$ on the surface $S\!=\!P_y$ in models A and B ($\varepsilon_x\!>\!0$) are shown in Figs. \ref{fig-5}(a) and (b), respectively. The difference is that $\vec{D}_{ij}\!=\!\vec{e}_{ij}$ in Fig. \ref{fig-5}(a) and $\vec{D}_{ij}\!=\!\vec{e}^{\;\prime}_{ij}$ in Fig. \ref{fig-5}(b) due to lattice deformation. The defined domains also differ. Figs. \ref{fig-6}(a)--(f) show $D^\mu_{S}$ in models A and B ($\varepsilon_x\!>\!0$) on $S\!=\!P_x$, $S\!=\!P_y$ and $S\!=\!P_z$.

The definition of the bulk DMI $D^\mu_V$ for models A and B also differs in ${e}_{ij}^{\mu}$ and ${e}_{ij}^{\prime\;\mu}$ such that
\begin{eqnarray}
\label{effective-D-SV-strain-AB}
\begin{split}
&D^\mu_V=\frac{1}{\sum_{ij\in V\setminus S}1}\sum_{ij\in V\setminus S} |{e}_{ij}^{\mu}|\;(\mu=x,y,z),\; S=P_x\cup P_y\cup P_z, \quad ({\rm model\; A}),\\
&D^\mu_V=\frac{1}{\sum_{ij\in V\setminus S}1}\sum_{ij\in V\setminus S} |{e}_{ij}^{\prime\;\mu}|\;(\mu=x,y,z),\; S=P_x\cup P_y\cup P_z, \quad ({\rm model\; B}).\\
\end{split}
\end{eqnarray}
 We note that the change from $\vec{e}_{ij}$ to $\vec{e}_{ij}^{\;\prime}$ when $\varepsilon_x\not\!=\!0$ effectively makes $D^\mu_{S, V}$ direction dependent, thus impacting magnetoelastic coupling. Therefore, neither uniaxial anisotropy, such as $-K_x \sum_i (\sigma_i^{x})^2$, nor more general magnetoelastic coupling terms are necessary in the Hamiltonian.

\begin{figure}[h!]
\centering{}\includegraphics[width=12.5cm,clip]{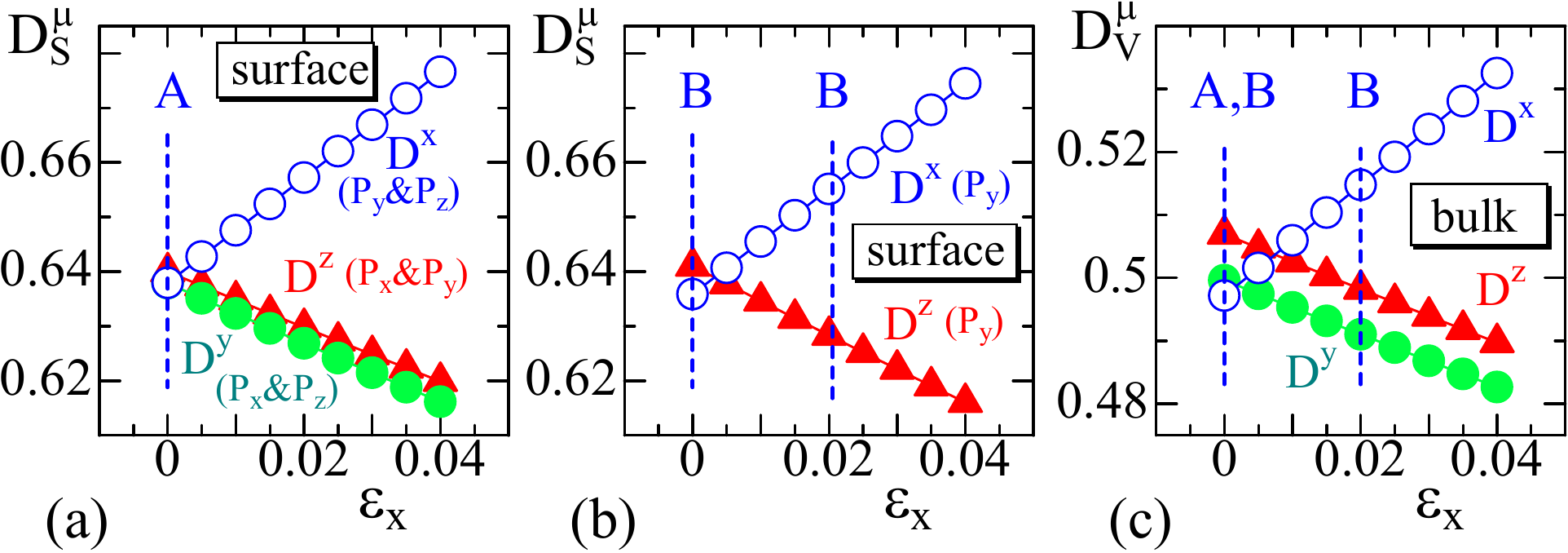}
\caption{The surface DMI $D^\mu_S (\mu\!=\!x, y, z)$ vs. $\varepsilon_x$ in (a) model A, (b) model B, and (c) the bulk DMI $D^\mu_V (\mu\!=\!x, y, z)$ vs. $\varepsilon_x$ in models A and B. The dashed lines with the symbols A and B in (a), (b) and (c) indicate the assumed strains $\varepsilon_x\!=\!0$ and $\varepsilon_x\!=\!0.02$ for the simulations in models A and B. The symbol $P_\mu\&P_\nu$ in (a) denotes $P_\mu\!\cup\! P_\nu$, where the corresponding surface DMI $D^*_S$ is obtained.
\label{fig-7} }
\end{figure}
$D^\mu_S$ vs. $\varepsilon_x$ and $D^\mu_V$ vs. $\varepsilon_x$ are plotted in Figs. \ref{fig-7}(a)--(c), where the lattice deformation defined by $\varepsilon_x$ is given in Eq. (\ref{lattice-deform}). The simulations of model A are performed only at $\varepsilon_x\!=\!0$, and the simulations of model B are performed at both $\varepsilon_x\!=\!0$ and $\varepsilon_x\!=\!0.02$, as indicated by the dashed lines in Figs. \ref{fig-7}(a)--(c). $D^\mu_S$ in Fig. \ref{fig-7}(a) is not always identical to $D^\mu_S$ in Fig. \ref{fig-7}(b) because the corresponding surfaces $S$ differ; however, we find that $D^\mu_S\to 0.64$  \cite{Koibuchi-etal-ICMsquare2022} for $\varepsilon_x\!\to\!0$ in both Figs. \ref{fig-7}(a) and (b).

Along the tensile strain direction $x$, $D^x_S$ and $D^x_V$ both increase with increasing strain $\varepsilon_x$. In Ref. \cite{YWang-etal-NatCom2020}, the corresponding coupling constant denoted by $D_{\rm ave}$ decreases with increasing tensile strain, and the response of $D_{\rm ave}$ to the tensile strain is opposite to that shown in the plotted data of $D^x_S$ and $D^x_V$ in Figs. \ref{fig-7}(a)--(c). This difference occurs because the sign of the DMI energy $S_{\rm DM}$ in Eq. (\ref{model-A}) and Eqs. (\ref{model-Fig-AB-1}), (\ref{model-Fig-AB-2}) is opposite to that in Ref. \cite{YWang-etal-NatCom2020} and is always negative because the continuous form of the DMI energy $\int \vec{\sigma}\cdot(\nabla\times\vec{\sigma})d^3x$ is replaced by the discrete expression $-\sum_{ij}\vec{e}_{ij}\cdot(\vec{\sigma}_i\times\vec{\sigma}_j)$. As a result, the changes in the two coupling constants $D^x_{S,V}$ and $D_{\rm ave}$ have the same effect on $S_{\rm DM}(<0)$ in model B and $S_{\rm DM}(>0)$ in the model in Ref. \cite{YWang-etal-NatCom2020}.

Note that without strain, $D^z_V$ is slightly larger than $0.5$ in Fig. \ref{fig-7}(c). This occurs because the distribution of $\vec{e}_{ij}$ deviates slightly from isotropic to nonisotropic in the $z$ direction. This type of anisotropy is expected in the case of tetrahedral lattices with flat boundary surfaces, where one side of each tetrahedron is forced to be on the same flat surface, and the area of $P_z$ is relatively large (Fig. \ref{fig-4}(a)). However, this deviation in $D^z_V$ in model B is constant, independent of the strain, and relatively small ($1.4\%$); hence, it does not have a substantial influence on the results.

\section{Simulation results\label{results}}
\subsection{Ground states and simulation details \label{sim-detail}}
Here, we comment on the ground states and the MC simulations. For the initial states in the MC simulations, the ground states are generated by the technique proposed in Ref. \cite {SEHog-etal-JMM2018}. This technique consists of minimizing the local energy of a given spin by (i) calculating the local field acting on it based on all of the terms in the Hamiltonian and (ii) aligning the spin in the direction of the local field. This approach minimizes the energy of the spin. Then, another spin is considered, and the same procedure is performed until all spins are considered. This constitutes a single step in the iterative process. Many iterations are performed until the system energy converges to a minimum. We generally use $10^6$ iterations for each run. To generate equilibrium configurations of the spin variables at a given temperature $T$ starting with the ground state, the Metropolis MC technique \cite{Metropolis-JCP-1953,Landau-PRB1976} is used. This procedure for updating the spin variables is the same as that described in Section \ref{DMI-dependent-conf}. The total number of MC sweeps (MCSs) is set to $1\times 10^8$ or $2\times 10^8$, with $2\times 10^8$ MCSs performed in the skyrmion phase and phase boundaries and $1\times 10^8$ MCSs performed in the other phases.

The simulation plan is as follows:
\begin{eqnarray}
  {\rm  model \;A, \; model \;B\; (\varepsilon_x=0)\; and \; model \; C \; \to \; model \;B\;  (\varepsilon_x>0)}.
\end{eqnarray}
First, simulations of model A and model B ($\varepsilon_x\!=\!0$) are performed to determine the difference between the results of the standard model ($\Leftrightarrow$ model A) and the results of the geometric confinement model ($\Leftrightarrow$ model B ($\varepsilon_x\!=\!0$)). In this first stage, simulations of model C are also performed, and the results are presented. Next, model B ($\varepsilon_x\!>\!0$) is simulated to observe the effects of both geometric confinement and uniaxial strains.
\subsection{Effect of zero DMI coefficients on the boundary surfaces \label{zero-DMI}}
\begin{figure}[th]
\centering{}\includegraphics[width=10.5cm,clip]{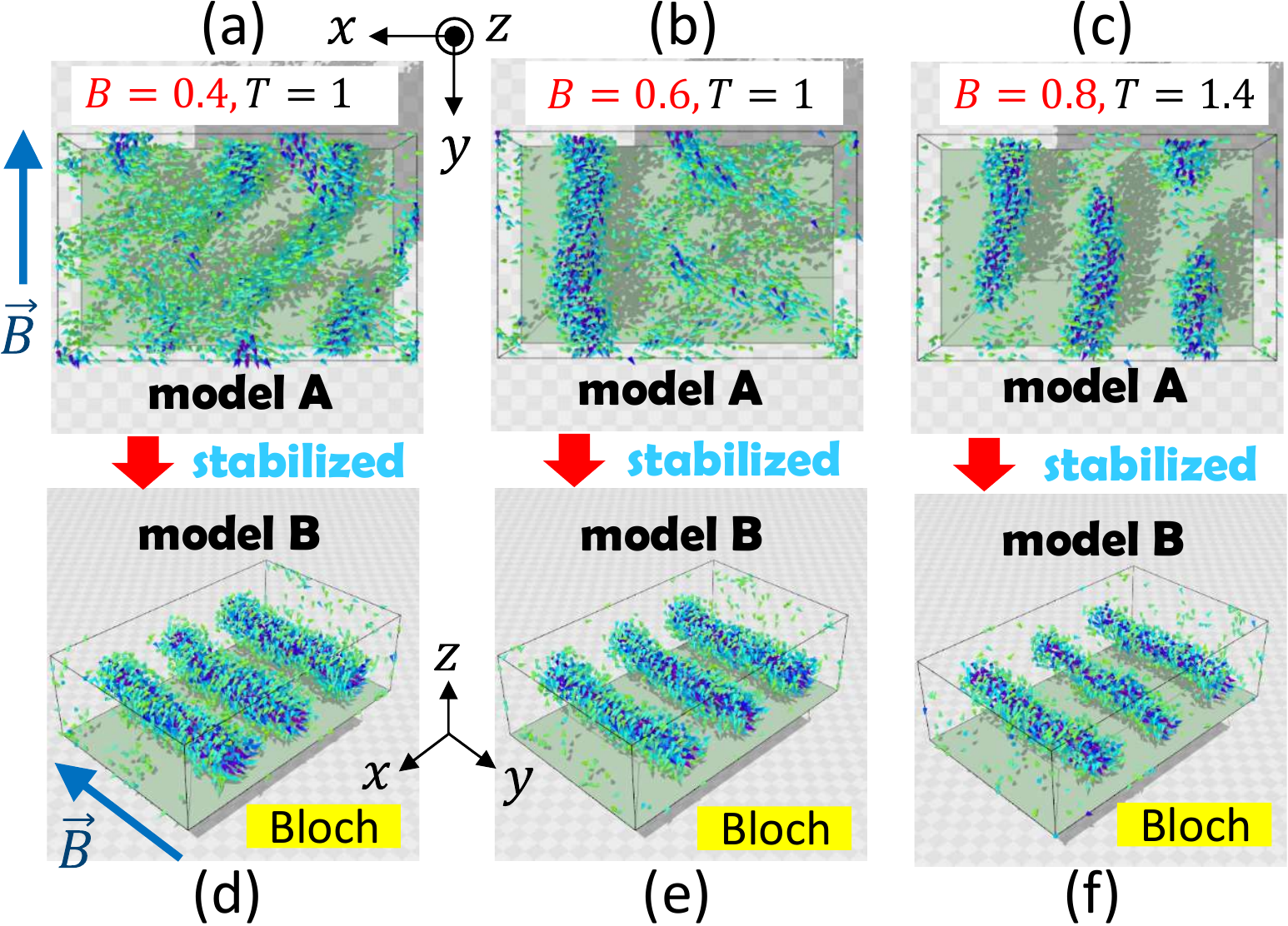}
\caption{Snapshots of Bloch-type skyrmions in model An obtained with the parameters $(B, T)$ of (a) $(0.4, 1)$, (b) $(0.6, 1)$, and (c) $(0.8, 1.4)$. (d), (e) and (f) Snapshots of model B
($\varepsilon_x\!=\!0$) obtained with the same parameters. The parameters $(\lambda, D)$ are fixed to $(\lambda, D)\!=\!(1, 0.9)$ in both models. The only difference between the models is the DMI coefficient on the boundary surfaces $P_x$ and $P_z$ described by $\Gamma_{ij}$ in Eq. (\ref{for-model-BC}). Skyrmions are visualized by showing only spins $\vec{\sigma}$ with $\sigma^y\geq 0$, where $\vec{B}\!=\!(0, -B, 0)$. Skyrmions are considerably stabilized in model B ($\varepsilon_x\!=\!0$) by surface effects. The diameter of the skyrmion strings decreases with increasing $B$, while the distance between the strings remains unchanged, as expected.
\label{fig-8} }
\end{figure}
First, we present snapshots of model A and model B ($\varepsilon_x\!=\!0$) in Figs. \ref{fig-8}(a)-(c) and Figs. \ref{fig-8}(d)-(f), respectively, to see how the zero DMI condition stabilizes skyrmions. The parameters are noted in the figures and captions. The snapshots in the upper row show that stable skyrmions do not occur in model An in these parameter regions, and the snapshots in the lower row show that these unstable skyrmions change to clearly separated stable skyrmions.

\begin{figure}[ht]
\centering{}\includegraphics[width=10.5cm,clip]{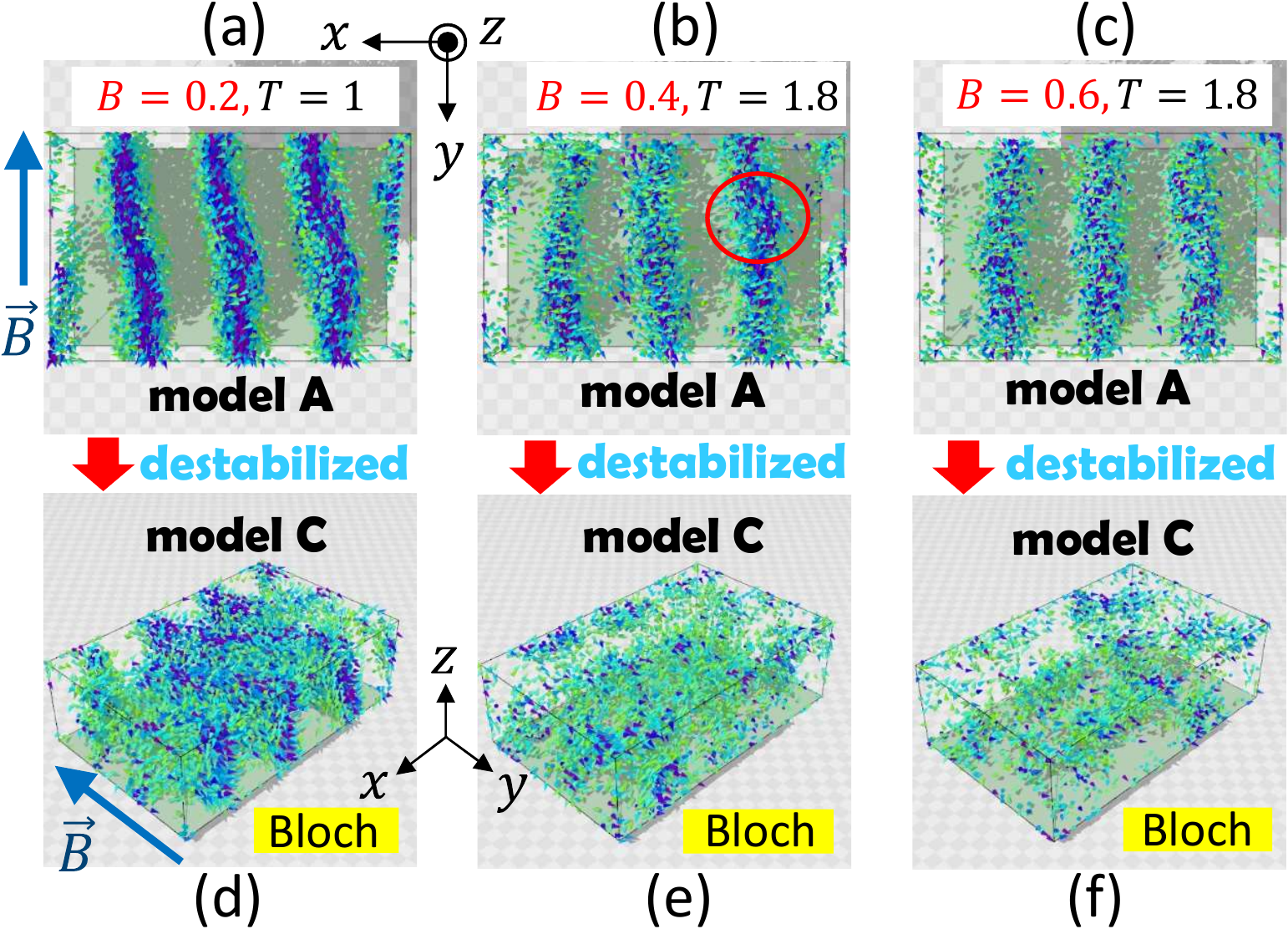}
\caption{
Snapshots of Bloch-type skyrmions in model A at $(B, T)$ of (a) $(0.2, 1)$ (nonconfined stripe), (b) $(0.4, 1.8)$ (nonconfined skyrmion: touching a surface (red circle)), and (c) $(0.6, 1.8)$ (confined skyrmion). (d), (e) and (f) show snapshots of model C obtained with the same parameters. The parameters $(\lambda, D)$ are fixed to $(1, 0.9)$ in both models. The only difference between the models is the FMI coefficient on the boundary surfaces $P_x$ and $P_z$ described by $\Gamma_{ij}$ in Eq. (\ref{for-model-BC}). The stripes in (a) and skyrmions in (b) and (c) are destabilized by surface effects in model C, in sharp contrast to the case of model B.
\label{fig-9} }
\end{figure}

We also compare the results of model A and model C, in which the zero FMI condition is assumed (Figs. \ref{fig-9}(a)--(c) and \ref{fig-9}(d)--(f)). The parameters $(\lambda, D)$ are fixed to $(\lambda, D)\!=\!(1, 0.9)$, which are the same as those in Fig. \ref{fig-8}, and the other parameters $(B, T)$ are shown in the figures. We find from the snapshots in Fig. \ref{fig-9} that configurations observed in model A, including skyrmion strings, become unstable in model C. Moreover, we checked that no confined skyrmion is observed by varying $(B, T)$ in the ranges $0.2\!\leq\!T\!\leq\!3.4$ and $0.2\!\leq\!B\!\leq\!1$. Thus, the heterogeneity of the FMI coefficient between the surfaces and bulk does not stabilize but rather destabilizes the skyrmions. Since no stabilization is observed in model C, we study model A, model B ($\varepsilon_x\!=\!0$) and model B ($\varepsilon_x\!>\!0$) henceforth.

\begin{figure}[h!]
\centering{}\includegraphics[width=10.5cm,clip]{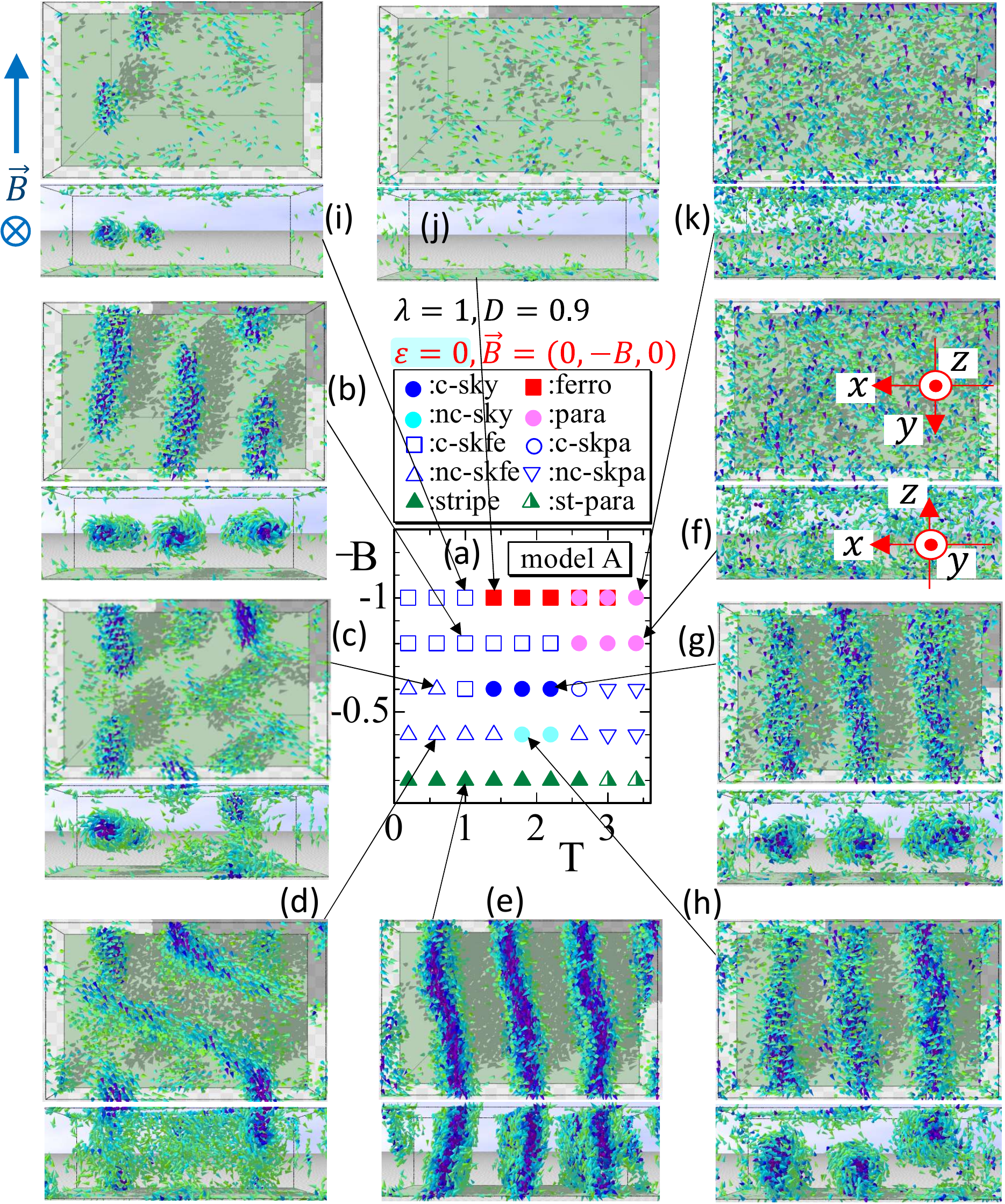}
\caption{(a) $BT$ phase diagram of model A, with (b)--(k) showing snapshots of the upper view (upper part) and side view (lower part). The upper views in (b) and (i) show that the skyrmions are incomplete and are denoted as the ``confined skyrmion ferromagnetic'' (c-skfe) phase. (j) ``ferro'': almost all spins of $\sigma^y\!<\!0$ are forced to the $\vec{B}$ direction; (c), (d) ``nc-skfe'': no complete skyrmion string is present, and some incomplete skyrmions touch the upper wall; (e) ``stripe'': three stripes touch the upper and lower walls; (f), (k) ``para'': the spin directions are approximately random; (g) ``c-sky'': three confined skyrmion strings are present; and (h) ``nc-sky'': three nonconfined skyrmion strings are present. See text for detailed description. \label{fig-10}    }
\end{figure}
To better understand the difference between model A and model B ($\varepsilon_x\!=\!0$), we plot $BT$ phase diagrams and snapshots of the two models in the ranges $0.2\leq B \leq 1.0$ and $0.2\leq T \leq 3.4$. The parameters $(\lambda, D)$ are fixed to $(\lambda, D)\!=\!(1, 0.9)$, similar to Figs. \ref{fig-8} and \ref{fig-9}. The $BT$ phase diagram of model A is plotted in Fig. \ref{fig-10}(a), and snapshots of this model are shown in Figs. \ref{fig-10}(b)--(k). Since our goal is not to precisely determine the phase boundary (which would require substantial computing time), the phase diagram is drawn by viewing the snapshots and determining whether the skyrmion configurations are stable during a sufficiently large number ($2\times\! 10^8$) of MCSs. In model A, the skyrmion phase can be divided into confined skyrmion (c-sky) and nonconfined skyrmion (nc-sky) phases, corresponding to Figs. \ref{fig-10}(g) and \ref{fig-10}(h), respectively. The nonconfined skyrmions in Fig. \ref{fig-10}(h) have oblong shapes along the $z$ direction, which is consistent with the experimental data presented in Refs. \cite{HDu-etal-NatCom2015,CJin-etal-NatCom2017}. The stripe phase is observed and can also be divided into confined and nonconfined phases. A nonconfined stripe configuration is shown in Fig. \ref{fig-10}(e), which is denoted as ``stripe''. Field-induced ferromagnetic (paramagnetic) configurations are expected to appear in the large-$B$ (high-$T$) region. The ``c-skfe'' snapshot in (i) represents an intermediate phase between the skyrmion and ferromagnetic phases and indicates that incomplete skyrmions are confined. Moreover, in (j), essentially all spins are $\sigma^y\!<\!0$, and hence, the configuration is denoted as ``ferro''. On the other hand, the snapshots in (k) and (f) obtained at $T\!=\!3.4$ show that the directions of essentially all spins are randomly distributed compared with those in (j), and therefore, the symbol ``para'' is used in (k) and (f). In addition, ``ferro'' and ``para'' are not always clearly separated, and therefore, both symbols corresponding to these two phases are used at points $(B, T)\!=\!(1, 2.6)$ and $(B, T)\!=\!(1, 3)$. Thus, the $BT$ phase diagram includes many symbols. Remarkably, the $BT$ phase diagram in Fig. \ref{fig-10}(a) shows that the confined skyrmion phase appears only in a small region in the central part of the diagram.

We note that these snapshots show the final configurations of the simulations of $1\!\times\!10^8\sim 2\!\times\!10^8$ MCSs starting with the ground state configurations, as mentioned in Section \ref{sim-detail}. Therefore, we consider that the obtained phase diagrams, 
including those presented below, are sufficiently stable. Note that our study is not focused on the orders of the phase transitions between the skyrmion phase and other phases.

\begin{figure}[h!]
\centering{}\includegraphics[width=10.5cm,clip]{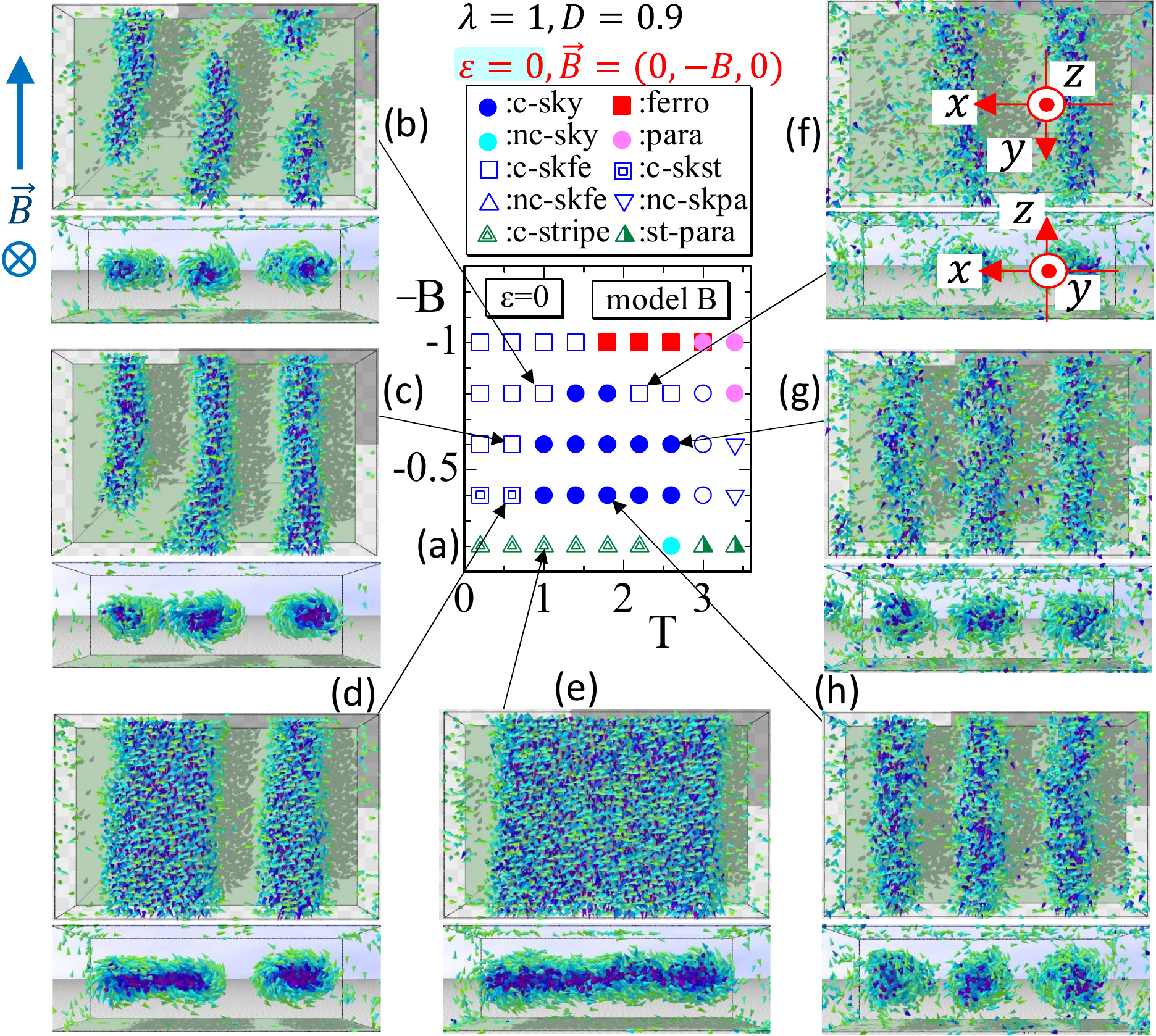}
\caption{(a) $BT$ phase diagram of model B ($\varepsilon_x\!=\!0$), with (b)--(h) showing snapshots of the spin configuration at several points. The area of the confined skyrmion phase (c-sky) in (a) is significantly larger than that in Fig. \ref{fig-10}(a). Skyrmion strings are not always complete in the confined sk-fe (c-skfe) phase in (b), (c), and (f), and three skyrmion strings that do not touch the walls are found in the c-sky phase in (g) and (h). A confined skyrmion stripe (c-skst) and confined stripe (c-stripe) appear in (d) and (e), respectively. The stripes in (d) and (e) are both in the $x$ direction, in contrast to the stripe configuration in Fig. \ref{fig-10}(e), which is in the $z$ direction.
\label{fig-11} }
\end{figure}
Next, we discuss the results obtained by model B with $\varepsilon_x\!=\!0$, as shown in Figs. \ref{fig-11}(a)--(h). The $BT$ phase diagram in Fig. \ref{fig-11}(a) shows that the area of the confined skyrmion phase is significantly larger than that in model A in Fig. \ref{fig-10}(a) in both the higher and lower regions on the $T$ axis and both directions on the $B$ axis. Moreover, the nonconfined phase, denoted by ``nc-$**$'', is observed only at $(B, T)\!=\!(0.2, 2.6), (0.4, 3.4), (0.6, 3.4)$ in this case. A comparison of the snapshots in Figs. \ref{fig-10}(d) and \ref{fig-11}(d), which are both obtained at $(B, T)\!=\!(0.4, 0.6)$, clearly shows that the skyrmion configurations are considerably stabilized due to the surface effects caused by $\Gamma_{ij}$ in Eq. (\ref{for-model-BC}). The nonconfined skyrmion states in Fig. \ref{fig-10}(h) also change to the confined skyrmion states in Fig. \ref{fig-11}(h), and the skyrmion shape changes from oblong to circular. This shape change occurs due to the surface effects introduced by $\Gamma_{ij}$. Additionally, we note that the stripe configurations in Fig. \ref{fig-10}(e) at $B\!=\!0.2$ change to confined stripe configurations, denoted by ``c-stripe'' in Fig. \ref{fig-11}(e). This change occurs due to the same surface effect.

\subsection{Stabilization by tensile strain
\label{tensile-strain}}
\begin{figure}[h!]
\centering{}\includegraphics[width=10.5cm,clip]{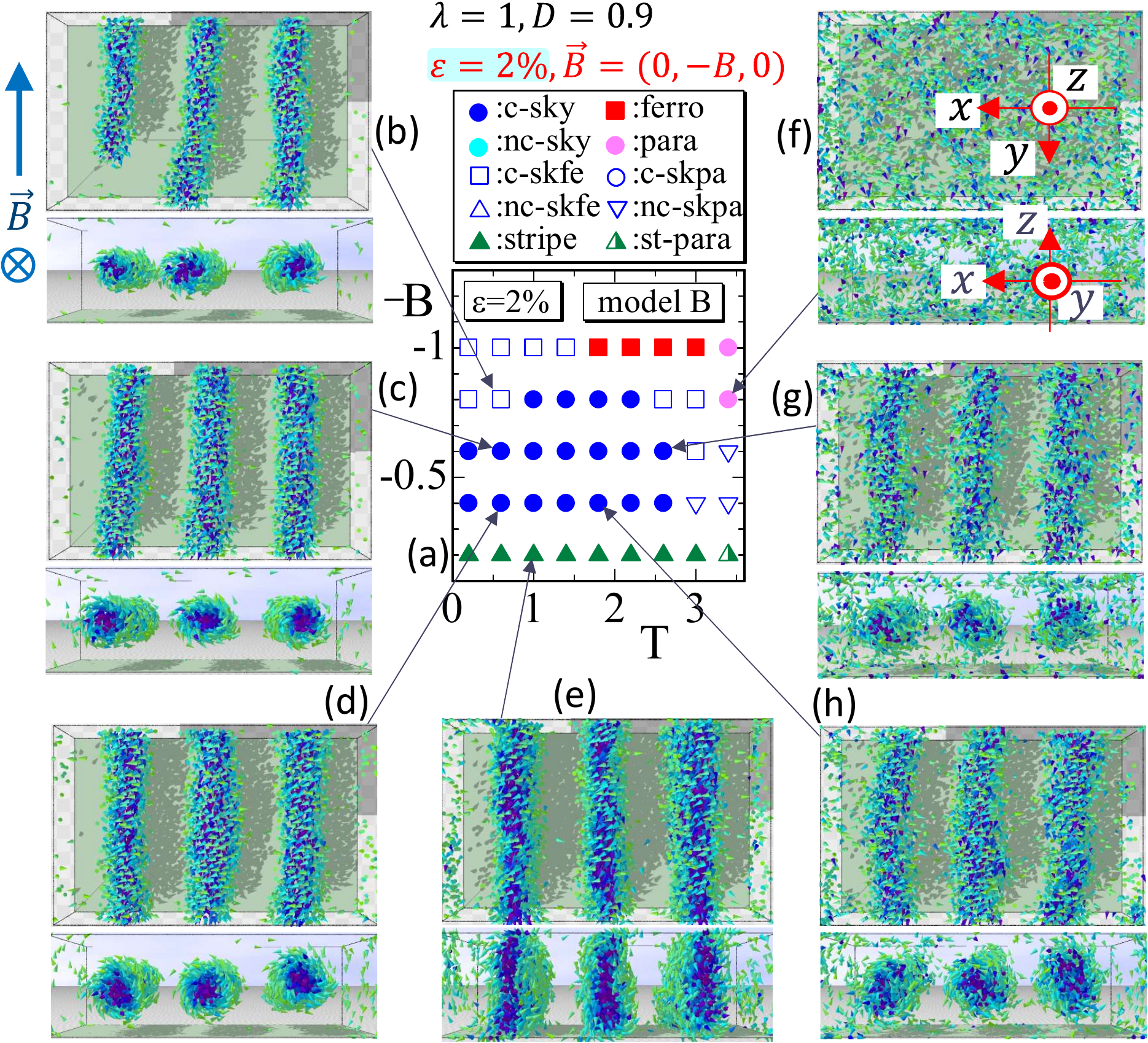}
\caption{(a) $BT$ phase diagram of model B ($\varepsilon_x\!>\!0$) under a tensile strain of $\varepsilon_x\!=\!0.02$, with (b)--(h) showing snapshots of the spin configuration at several points. The area of the confined skyrmion phase (c-sky) in (a) is larger than that in Fig. \ref{fig-11}(a). Skyrmion strings are complete in (c) and (d), in contrast to those in Figs. \ref{fig-11}(c) and (d). Some of the stripes in (e) do not touch the upper and lower boundaries and are confined, and the stripe direction also changes to vertical from horizontal in Fig. \ref{fig-11}(e). These changes are caused by the combined effect of the strains and GC.
\label{fig-12}  }
\end{figure}
In this subsection, we discuss the results of model B ($\varepsilon_x\!>\!0$) obtained under a small strain of $\varepsilon_x\!=\!L_x/L_x^0\!-\!1\!=\!0.02$ along the $x$ axis (see Fig. \ref{fig-4}(b) and Eq. (\ref{lattice-deform})). The $BT$ phase diagram and snapshots are shown in Figs. \ref{fig-12}(a)--(h). In this case, a nonconfined skyrmion phase does not appear. Moreover, the confined skyrmion phases at $B\!=\!0.4$ and $B\!=\!0.6$ extend to the low $T$ region, including the lowest temperature of $T\!=\!0.2$. The stripe in Fig. \ref{fig-12}(e) is along the $z$ direction, which differs from the stripe along the $y$ direction observed in Fig. \ref{fig-11}(e), and the stripe returns to the same direction as that in model A, as shown in Fig. \ref{fig-10}(e).
We note that the stripes in Fig. \ref{fig-12}(e) are partly or almost entirely confined, and this behavior differs from that of the stripes in Fig. \ref{fig-10}(e), which are not confined.
These changes in Fig. \ref{fig-12}(e) relative to the results presented in Figs. \ref{fig-11}(e) and \ref{fig-10}(e) are caused by the combined effect of strains and GC.

\subsection{Effect of DMI anisotropy caused by strains \label{DMI-anisotropy}}
\begin{figure}[h!]
\centering{}\includegraphics[width=9.5cm,clip]{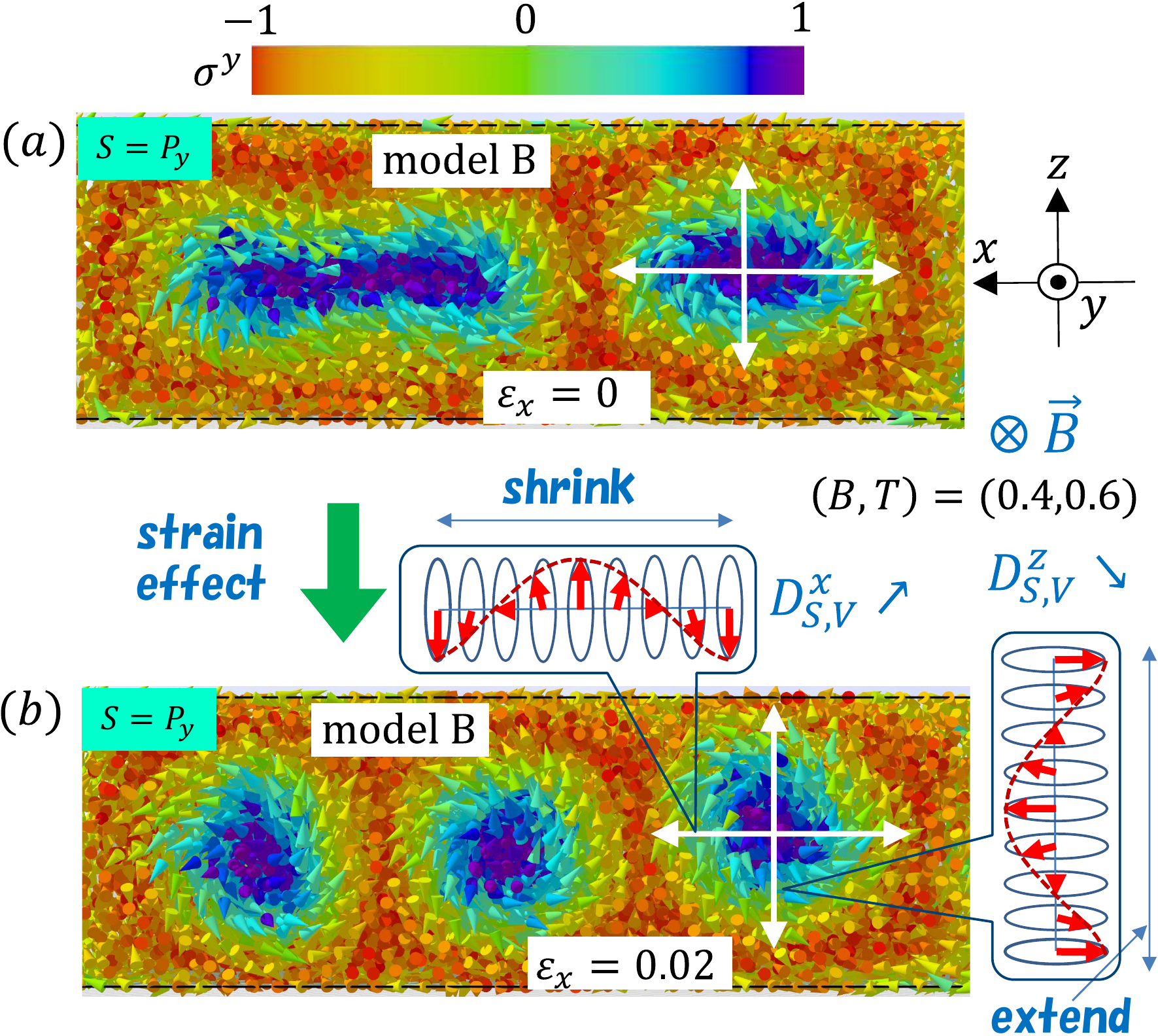}
\caption{Illustration of a tensile strain effect caused by lattice deformation along the $x$ axis in model B, with snapshots of (a) $\varepsilon_x\!=\!0$ and (b) $\varepsilon_x\!=\!0.02$, where the surface DMI is assumed to be $D^\mu_S\!=\!0$ on the boundaries $P_x$ and $P_z$ (Fig. \ref{fig-4}(a)). An increase in $D^x_{S, V}$, denoted by $D^x_{S, V}\!\!\nearrow$, effectively enlarges $|\vec{\sigma}_i\!\times\!\vec{\sigma}_j|$ in $S_{\rm DM}$, causing the skyrmion diameter to decrease along the $x$ axis, while a decrease in $D^z_{S, V}$, denoted by $D^z_{S, V}\!\!\searrow$, causes the skyrmion diameter to increase along the $z$ axis.\label{fig-13} }
\end{figure}
In this subsection, we discuss the role of DMI anisotropy in skyrmion stabilization in detail. The increases/decreases in $D^x_{V}$ and $D^z_{V}$ plotted in Figs. \ref{fig-7}(a)--(c) are caused by tensile strains with $\varepsilon_x(>\!0)$, and these variations in $D^x_{V}$ and $D^z_{V}$ influence the skyrmion configurations in model B. The skyrmion strings along the $y$ direction are influenced by $D^x_{V}$ and $D^z_{V}$ because the diameter ratio of the string depends on the characteristic lengths $2\lambda/ D^x_{V}$ and $2\lambda/D^z_{V}$ \cite{Butenko-etal-PRB2010}. Note that $D^\mu_{S}\!=\!0$ on $P_x$ and $P_z$ in model B. Therefore, if zero DMI coefficients are assumed on $P_y$, the characteristic lengths diverge, implying that well-defined skyrmion configurations are not expected on $P_y$. This motivates us to assume zero DMI coefficients only on $P_x$ and $P_z$ in model B (Fig. \ref{fig-4}(a)). 
In addition, the zero DMI coefficients on $P_x$ and $P_z$ suggest that DMIs and FMIs do not compete, preventing nonconfined skyrmions from appearing on $P_x$ and $P_z$. 
The snapshots in Figs. \ref{fig-13}(a) and (b) correspond to those in Fig. \ref{fig-11}(d) for $\varepsilon_x\!=\!0$ and Fig. \ref{fig-12}(d) for $\varepsilon_x\!=\!0.02$, respectively, where $(B, T)\!=\!(0.4, 0.6)$. The snapshots are drawn using all spins, in contrast to those in Figs. \ref{fig-8}-\ref{fig-12}, where only spins with $\sigma^y\!\geq\! 0$ are plotted. Figs. \ref{fig-13}(a) and (b) show that skyrmion configurations are not of meron or bimeron nature, in which the spin direction changes from $\sigma^y\!=\!1$ at the center to $\sigma^y\!=\!0$ at the periphery \cite{Zhang-etal-JPhys2020}.

These snapshots in Figs. \ref{fig-13}(a) and (b) show the effect of the nonzero strain $\varepsilon_x(=\!0.02)$. We emphasize that confined skyrmions can exist only in the central region between the plates $P_z$ because the zero DMI coefficients ($\Leftrightarrow\!D^\mu_{S}\!=\!0$) on $P_x$ and $P_z$ prevent skyrmion configurations on these surfaces. Thus, the confined skyrmions effectively feel repulsion from $P_z$, which has a width that is either not much larger than or comparable to the skyrmion size. As a result, the stripe configurations become parallel to $P_z$ in this region of $B$ and $T$, as shown in Fig. \ref{fig-13}(a). Moreover, these anisotropic stripe configurations change to skyrmion configurations and are stabilized by the variations in $D^x_{V}$ and $D^z_{V}$ due to the strain effect, as shown in Fig. \ref{fig-13}(b). If the width $L^z_0$ of the plate becomes sufficiently large, the stripe direction tends to be isotropic and not always parallel to $P_z$; therefore, the tensile strain along the $x$ direction is not always effective for stabilization.

Furthermore, we note that the oblong shape of the skyrmions along the $x$ direction (Fig. \ref{fig-13}(a)) in the low-magnetic-field region ($\Leftrightarrow\!B\!=\!0.4$) differs from that in the FeGe nanostripes in Ref. \cite{HDu-etal-NatCom2015}, where shape deformation is observed in the $z$ direction as the stripe width increases. However, this oblong shape along the $x$ direction is also observed in the same material, namely, FeGe, when the width is sufficiently narrow \cite{CJin-etal-NatCom2017}. Therefore, the result shown in Fig. \ref{fig-13}(a) is consistent with the results in Refs. \cite{HDu-etal-NatCom2015,CJin-etal-NatCom2017}. On the other hand, the response of the skyrmion shape to stresses with respect to the stability in ${\rm Cu_2OSeO_3}$ differs from those in FeGe and MnSi because skyrmions are stabilized in ${\rm Cu_2OSeO_3}$ if tension is applied perpendicular to the $\vec{B}$ axis \cite{Seki-etal-PRB2017}, while in MnSi in Ref. \cite{Nii-etal-NatCom2015,Charcon-etal-PRL2015}, stabilization occurs when compression is applied perpendicular to the $\vec{B}$ axis. Thus, our model B results are consistent with the skyrmion response in ${\rm Cu_2OSeO_3}$ in Ref. \cite{Seki-etal-PRB2017}. Since strain and GC effects are both implemented in model B, these findings indicate that the same GC effects in sufficiently narrow nanostripes occur in different materials, with the strain effects depending on the material. Here, both effects modify the effective couplings $D^y_{V}$ and $D^z_{V}$ in model B, implying that the strain-induced variations in $D^y_{V}$ and $D^z_{V}$ are material-dependent even though the changes in the skyrmion shape due to the direction-dependent DMI coefficients are independent of the material.
The changes in the direction-dependent DMI coefficients in response to the strain and the skyrmion morphology due to variations in the nanostripe width are both interesting.
However, the numerical data presented in this paper are insufficient for studying these problems, and further numerical studies are necessary.

Finally, in this subsection, to confirm that the skyrmion phases plotted in Figs. \ref{fig-10}(a), \ref{fig-11}(a) and \ref{fig-12}(a) are reasonable, we calculate the topological charge $N_{\rm sk}$ corresponding to the total number of skyrmions, which is defined as
\begin{eqnarray}
\label{top-charge}
N_{\rm sk}=\frac{1}{4\pi}\int d^2x \vec{\sigma}\cdot \frac{\partial \vec{\sigma}}{\partial x_1}\times \frac{ \partial \vec{\sigma}}{\partial x_2}
\end{eqnarray}
on the surface $P_y$ on the side with the maximum $y$ (see Figs. \ref{fig-2}(a), (b)), which is not shown in Fig. \ref{fig-4}(a). The local coordinates $x_1, x_2$ on the right-hand side of Eq. (\ref{top-charge}) are defined along the axes of the triangles on the surface $P_y$ (Appendix \ref{App-B}). The differentials ${\partial \vec{\sigma}}/{\partial x_i}, (i\!=\!1,2)$ are numerically evaluated according to the differences
(see Appendix \ref{App-B} for the discrete form of $N_{\rm sk}$), and therefore, the calculated results are continuous and not the same as the integer variations visually observed using the snapshots in Figs. \ref{fig-10}--\ref{fig-12}. In addition, $N_{\rm sk}$ is calculated using only configurations on $P_y$, and therefore, this value does not always reflect information about the skyrmion strings inside the 3D tetrahedral lattice in Fig. \ref{fig-4}(a). However, we expect that the curves of $N_{\rm sk}$ vs. $T$ should reflect how skyrmions are influenced by thermal fluctuations.
\begin{figure}[h]
\centering{}\includegraphics[width=10.5cm,clip]{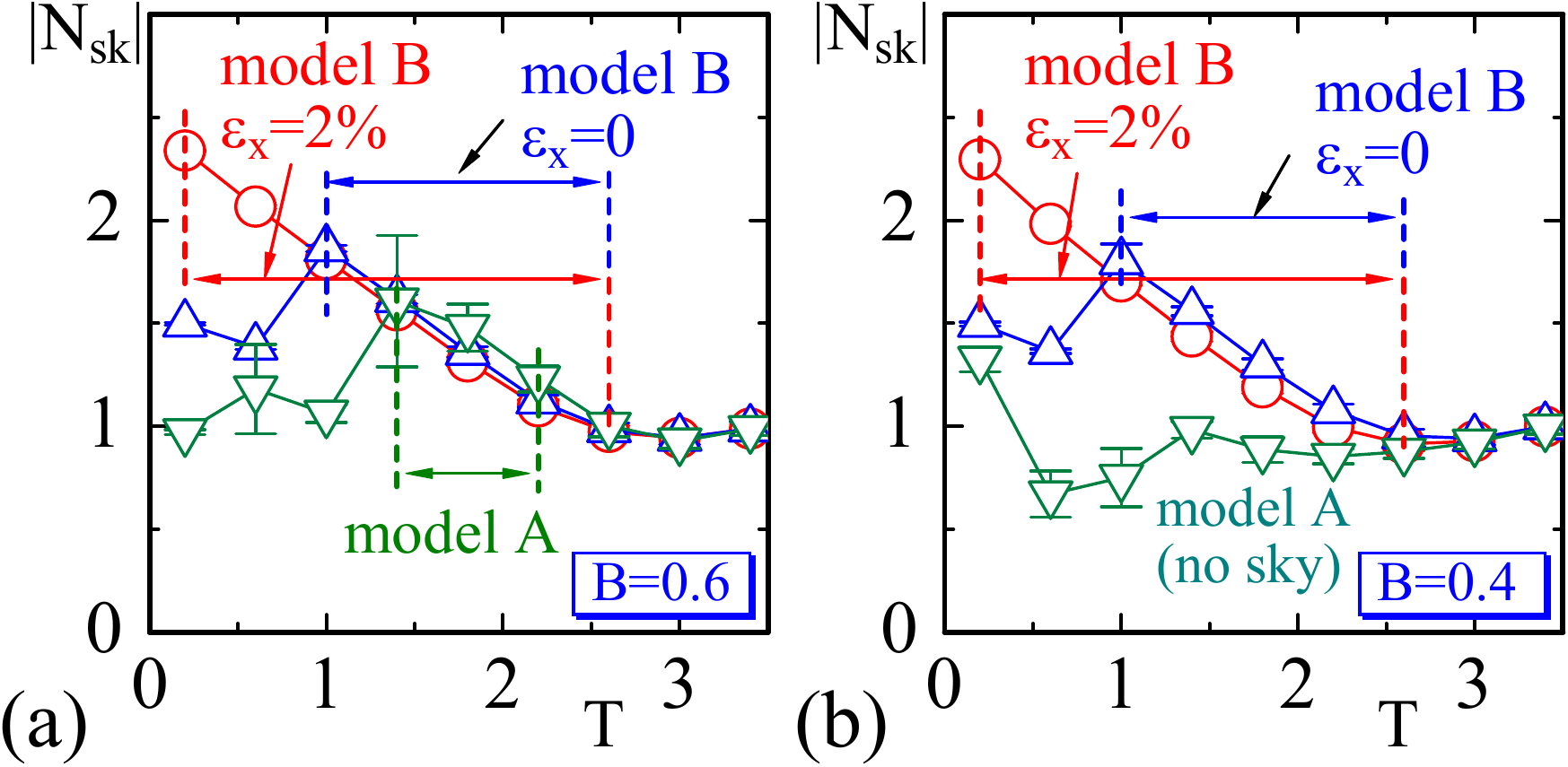}
\caption{Topological charge $|N_{\rm sk}|$ vs. $T$ obtained under (a) $B\!=\!0.6$ and (b) $B\!=\!0.4$ for model A (\textcolor{green}{$\bigtriangledown$}), model B with $\varepsilon_x\!=\!0$ (\textcolor{blue}{$\bigtriangleup$}) and model B with $\varepsilon_x\!=\!2\%$ (\textcolor{red}{$\bigcirc$}), corresponding to the data in the phase diagrams in Figs. \ref{fig-10}(a), \ref{fig-11}(a), and \ref{fig-12}(a), respectively. The dashed vertical lines indicate the temperature region of the skyrmion phase in each curve, and $N_{\rm sk}$ is meaningful only in these regions.
\label{fig-14} }
\end{figure}

The absolute values $|N_{\rm sk}|$ obtained under $B\!=\!0.6$ and $B\!=\!0.4$ are plotted in Figs. \ref{fig-14}(a) and (b), respectively. The sample configurations for the calculation of $|N_{\rm sk}|$ are obtained every $1000$ MCSs during the $1\times 10^8 \sim 2\times 10^8$ MCSs, as described in Section \ref{sim-detail}. The error bar denotes the standard error obtained by the binning analyses in the MC simulations \cite{Janke-2002}. The large error bar on the data of model A indicates the relatively large fluctuation in $|N_{\rm sk}|$, implying that the skyrmion phase is not always stable. The dashed vertical lines denote the skyrmion regions in $T$ corresponding to the data (\textcolor{blue}{$\bullet$}) in the phase diagrams of Figs. \ref{fig-10}(a), \ref{fig-11}(a), and \ref{fig-12}(a). The decrease in $|N_{\rm sk}|$ with increasing $T$ in the skyrmion phases in Figs. \ref{fig-14}(a) and (b) implies that thermal fluctuations influence the skyrmion shape.

Moreover, we find that $|N_{\rm sk}|$ is essentially independent in model A, model B ($\varepsilon_x\!=\!0$) and model B ($\varepsilon_x\!=\!2\%$), at least in the skyrmion region with $B\!=\!0.6$ and $B\!=\!0.4$. The $|N_{\rm sk}|$ values of model A (\textcolor{green}{$\bigtriangledown$}) in Fig. \ref{fig-14}(b) are clearly different from those of model B, indicating that the numerically calculated $N_{\rm sk}$ correctly reflects the topological charge of the skyrmions.
Furthermore, $|N_{\rm sk}|$ decreases discontinuously in the low $T$ region at the phase boundary between the skyrmion and other phases. More specifically, $|N_{\rm sk}|$ in model B with zero strain $\varepsilon_x\!=\!0$ discontinuously changes at $T\!\to\!1$ in Fig. \ref{fig-14}(a) for $B\!=\!0.6$ and Fig. \ref{fig-14}(b) for $B\!=\!0.4$. In addition, $|N_{\rm sk}|$ in model A in Fig. \ref{fig-14}(a) discontinuously changes at $T\!\to\!1.4$.
 These discontinuities are consistent with the visually observed phase boundaries between the c-sky and c-skfe phases in the phase diagrams presented in Figs. \ref{fig-11}(a) and \ref{fig-10}(a). In contrast, $|N_{\rm sk}|$ in model B with $\varepsilon_x\!=\!2\%$ has no discontinuities in the skyrmion region $0.2\!\leq\!T\!\leq\!2.6$ for both $B\!=\!0.6$ and $B\!=\!0.4$. This smooth variation implies that the skyrmion phase in model B with $\varepsilon_x\!=\!2\%$ is stable. The reason why $|N_{\rm sk}|$ increases with decreasing $T$ is that the skyrmion configuration is more stable in the low-temperature region, as shown by comparing the snapshots in Fig. \ref{fig-12}(d) and Fig. \ref{fig-12}(h) at $B\!=\!0.4$. For such stable skyrmion configurations, the discrete expression in Eq. (\ref{discrete-Nsk}) is relatively accurate. In contrast, for fluctuating skyrmion configurations, evaluating $|N_{\rm sk}|$ with a discrete expression is less accurate, as previously mentioned. Thus, we consider that $|N_{\rm sk}|$ decreases with increasing $T$ in the region $0.2\!\leq\!T\!\leq\!2.6$.

Importantly, these relatively rapid variations in $|N_{\rm sk}|$ with respect to $T$ in the range $0.2\!\leq\!T\!\leq\!2.6$ occur because $|N_{\rm sk}|$ is very small, with $|N_{\rm sk}|\!=\!3$ for a small region of $T$. If the lattice is sufficiently large, the slopes of the curves are expected to be moderate. In this sense, the results plotted in Figs. \ref{fig-14}(a) and (b) depend on the lattice size. We note that the surface effect on confinement implemented by $\Gamma_{ij}\!=\!0$ is expected to be weak if the thickness of the lattice is much larger than the skyrmion size. In this paper, as described in the first part of Section \ref{GC-model}, we assume that the lattice is sufficiently thin and approximately twice as large as the skyrmion size to clarify the surface effects, as shown in the snapshots in Fig. \ref{fig-12}. The specific value of the lattice thickness at which the surface effects disappear is interesting; however, this problem remains to be studied in the future. 

\section{Summary and conclusion\label{conclusion} }
In this paper, we numerically study skyrmion stabilization using a plate-shaped 3D lattice discretized by tetrahedra by assuming zero Dzyaloshinskii-Moriya interaction (DMI) coefficients on the boundary surfaces parallel to the magnetic field to evaluate geometric confinement (GC) effects. The Hamiltonian is given by a linear combination of the standard ferromagnetic interaction (FMI) energy, the DMI energy for Bloch-type skyrmions and the Zeeman energy.

Compared with the nonzero surface DMI model, the stabilization effect is significantly improved in the zero surface DMI model, with an increase in the area of the skyrmion phase in the $BT$ phase diagram.
Moreover, the tensile strain implemented by lattice deformation enhances skyrmion stabilization, extending the skyrmion phase in the low-temperature region. This strain-induced enhancement is observed only in the model with zero DMI coefficients on the surface, where zero DMI conditions are implemented as a GC effect. The numerical data indicate that the zero DMI condition on the surface competes with tensile strain, thereby enhancing the skyrmion phase stability.
In addition, we verified that stability is not always observed in a model with zero FMI coefficients on the boundary surfaces. This observation supports that the zero DMI condition model is meaningful as a GC model.

The models in this paper are applicable to the skyrmions observed in ${\rm Cu_2OSeO_3}$, in which tensile strains perpendicular to the magnetic field stabilize the skyrmions. Moreover, the mechanism by which anisotropic DMI coefficients stabilize the skyrmions in ${\rm Cu_2OSeO_3}$ is expected to be similar to that for skyrmion stabilization in MnSi and FeGe because the variations in the skyrmion shape according to the anisotropic DMI coefficients should be the same. However, detailed information regarding the shape morphology of the confined skyrmions and the dependence on the domain size has not yet been obtained in the framework of effective interaction theories such as Finsler geometry models implementing FMI and DMI anisotropy to assess the effects of external stimuli and GC.
Therefore, additional theoretical and numerical studies are necessary to develop a unified understanding of the stability leading to skyrmion control.

\acknowledgements
This work was supported in part by a JSPS Grant-in-Aid for Scientific Research (19KK0095) and Collaborative Research Project J20Ly18 at the Institute of Fluid Science (IFS), Tohoku University. The numerical simulations were performed on the supercomputer system AFI-NITY at the Advanced Fluid Information Research Center, IFS, Tohoku University.

\appendix

\section{Construction of a 3D lattice by tetrahedrons \label{App-A}}
\begin{figure}[h]
\centering{}\includegraphics[width=9.5cm,clip]{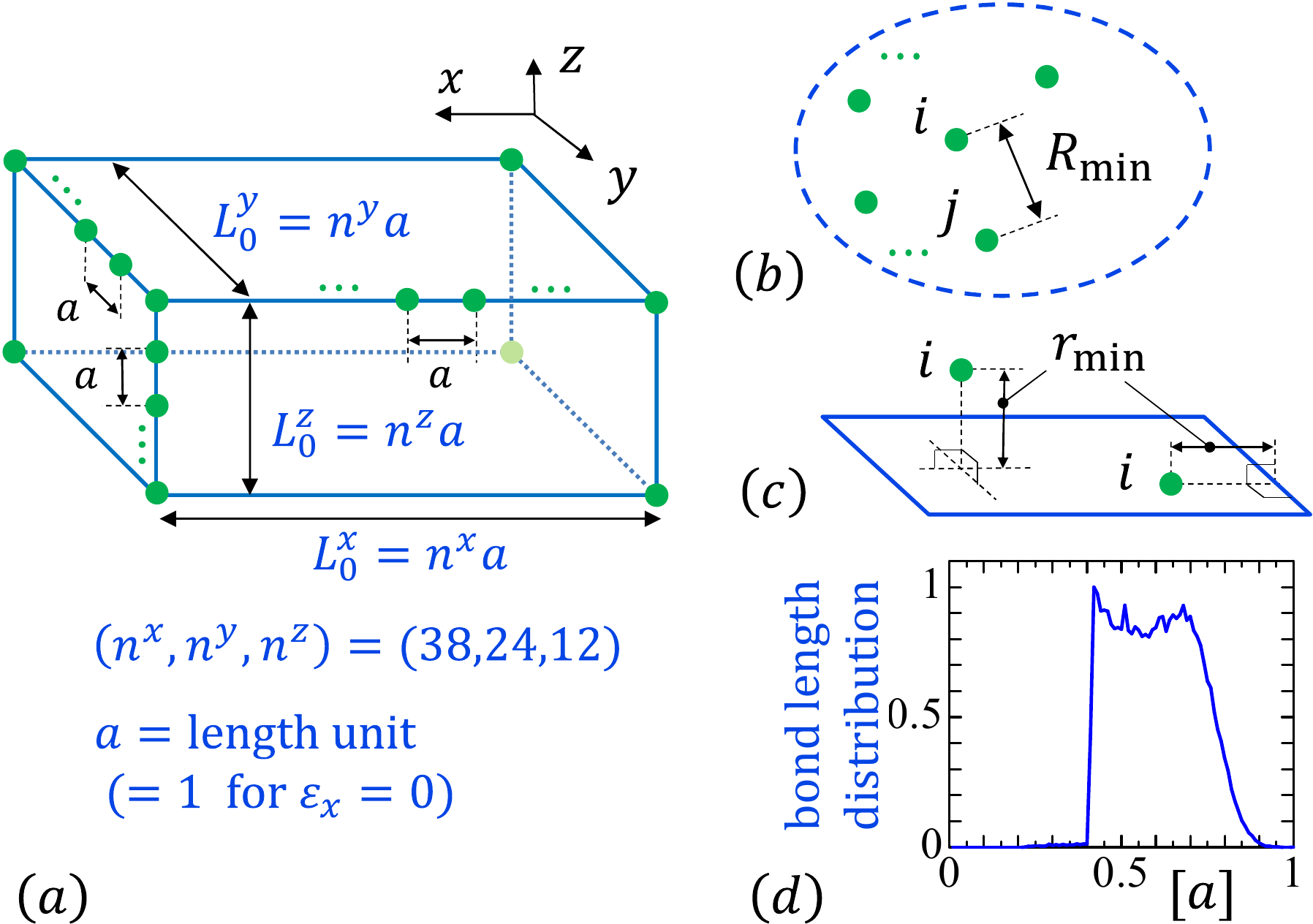}
\caption{
(a) An illustration of a 3D cubic lattice, the shape of which is characterized by $(n^x, n^y, n^z)\!=\!(38, 24, 12)$. (b) The minimum distance $R_{\rm min}$ between two vertices $i$ and $j$ for the bond length inside and between $i$ and $j$ on the surfaces is $R_{\rm min}\!=\!0.8a$, (c) the minimum distance $r_{\rm min}$ between vertex $i$ inside and the surfaces is given by $r_{\rm min}\!=\!0.43a$, and the minimum distance between surface vertex $i$ and the edges is also given by the same $r_{\rm min}$.
In addition to these constraints, small random numbers are used to move the vertex position to link the vertices by Voronoi tessellation \cite{Friedberg-Ren-NPB1984}. (d) The normalized distribution of the bond lengths.
\label{fig-15} }
\end{figure}
We briefly present the construction of the 3D lattice on which the models are defined. The edge length of the cube along the $\mu(=x, y, z)$ direction is given by $n^\mu a$, with $(n^x, n^y, n^z)\!=\!(38, 24, 12)$, while the total number of vertices on the edge is $n^\mu\!+\!1$. The length unit or the lattice spacing $a$ can be fixed at an arbitrary number, and hence, $a\!=\!1$ for zero strain $\varepsilon_x\!=\!0$.
The vertices on the edges parallel to the $\mu (=x, y, z)$ axis are separated by $a$, and the edge length is given by $n^\mu$. The vertices inside and on the surfaces are randomly distributed with a minimum distance $R_{\rm min}(=\!0.8a)$ (Fig. \ref{fig-15}(b)) and a minimum distance $r_{\rm min}(=\!0.43a)$ from the surfaces (Fig. \ref{fig-15}(c)). The surface vertices are separated by the same minimum distance $r_{\rm min}$ from the edges. The positions of the vertices except those on the edges fluctuate with additional small random numbers. The Voronoi tessellation technique is used to link the vertices \cite{Friedberg-Ren-NPB1984}, and the bond lengths $\ell$ are distributed mainly in the range $0.4a\!\leq\!\ell\!\leq\! 0.8a$ (Fig. \ref{fig-15}(d)).

The lattice size is given by $(N, N_{\rm B}, N_{\rm T}, N_{\rm tet})\!=\!(14548, 100313, 167930, 82164)$, where $N, N_{\rm B}, N_{\rm T}$ and $N_{\rm tet}$ are the total numbers of vertices, bonds, triangles, and tetrahedra. These numbers satisfy the condition $N\!-\!N_{\rm B}\!+\!N_{\rm T}\!-\!N_{\rm tet}\!=\!1$, which is the same condition as in tetrahedron $(N, N_{\rm B}, N_{\rm T}, N_{\rm tet})\!=\!(4, 6, 4, 1)$.

\section{
Discrete form of the topological charge \label{App-B}}
\begin{figure}[h]
\centering{}\includegraphics[width=10.5cm,clip]{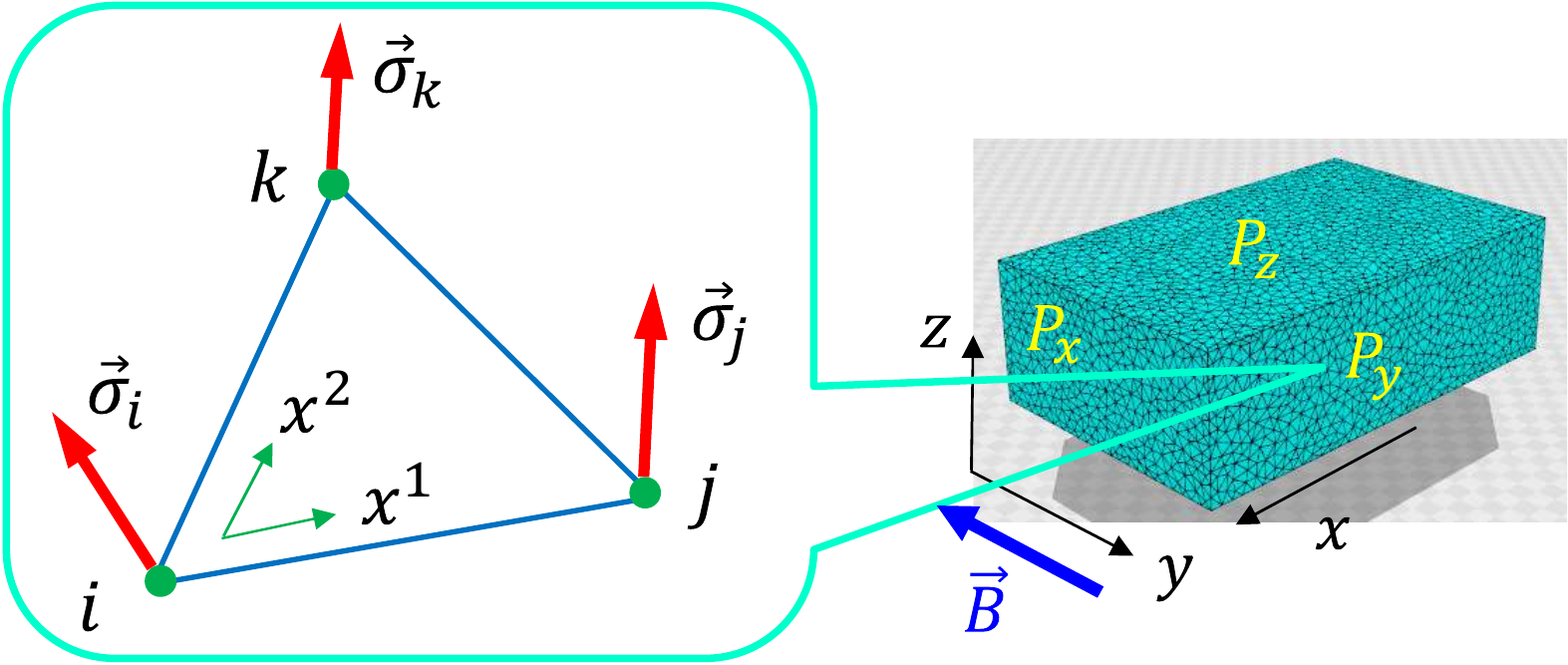}
\caption{A local coordinate $(x^1, x^2)$ and spin configurations $\vec{\sigma}_i$, $\vec{\sigma}_j$ and $\vec{\sigma}_k$ on triangle ${\it \Delta}_{ijk}$ on surface $P_y$. On this surface $P_y$, the topological charge $N_{\rm sk}$ in Eq. (\ref{top-charge}) is calculated by using the discrete expression of $N_{\rm sk}$ in Eq. (\ref{discrete-Nsk}).
\label{fig-16} }
\end{figure}
We present a discrete form of $N_{\rm sk}\!=\!\frac{1}{4\pi}\int d^2x \vec{\sigma}\cdot \frac{\partial \vec{\sigma}}{\partial x_1}\times \frac{ \partial \vec{\sigma}}{\partial x_2}$ in Eq. (\ref{top-charge}) in this Appendix. First, the integral $\int d^2x$ is replaced by a sum over triangles $\sum_{{\it \Delta}\in P_y}$ on surface $P_y$ on one side (Fig. \ref{fig-16}). On a triangle ${\it \Delta}_{ijk}$ with vertices $i, j$ and $k$, the differentials are replaced by ${\partial \vec{\sigma}}/{\partial x_1}\!\to\! \vec{\sigma}_j\!-\!\vec{\sigma}_i$ and ${\partial \vec{\sigma}}/{\partial x_2}\!\to\! \vec{\sigma}_k\!-\!\vec{\sigma}_i$, where $(x^1, x^2)$ is a local coordinate of the triangle ${\it \Delta}_{ijk}$. Therefore, we obtain $\vec{\sigma}\cdot \frac{\partial \vec{\sigma}}{\partial x_1}\!\times\! \frac{ \partial \vec{\sigma}}{\partial x_2}\to \vec{\sigma}_i\cdot \vec{\sigma}_j\!\times\!\vec{\sigma}_k$ on ${\it \Delta}_{ijk}$. On triangle ${\it \Delta}_{ijk}$, we have two other local coordinate origins at vertices $j$ and $k$. Therefore, by including $ \vec{\sigma}_j\cdot \vec{\sigma}_k\!\times\!\vec{\sigma}_i$ and $\vec{\sigma}_k\cdot \vec{\sigma}_i\!\times\!\vec{\sigma}_j$ with the factor $1/3$, we obtain the replacement $\vec{\sigma}\cdot \frac{\partial \vec{\sigma}}{\partial x_1}\!\times\! \frac{ \partial \vec{\sigma}}{\partial x_2}\to (1/3)[\vec{\sigma}_i\cdot \vec{\sigma}_j\!\times\!\vec{\sigma}_k\!+\!\vec{\sigma}_j\cdot \vec{\sigma}_k\!\times\!\vec{\sigma}_i\!+\!\vec{\sigma}_k\cdot \vec{\sigma}_i\!\times\!\vec{\sigma}_j]$ on ${\it \Delta}_{ijk}$. Thus, the discrete form of $N_{\rm sk}$ is given by
 \begin{eqnarray}
 \label{discrete-Nsk}
N_{\rm sk} =\frac{1}{12\pi}\sum_{{\it \Delta}_{ijk}\in P_y}
 \left[\vec{\sigma}_i\cdot (\vec{\sigma}_j\times\vec{\sigma}_k)+\vec{\sigma}_j\cdot (\vec{\sigma}_k\times\vec{\sigma}_i)+\vec{\sigma}_k\cdot (\vec{\sigma}_i\times\vec{\sigma}_j )\right].
\end{eqnarray}

\section*{References}


\end{document}